\documentclass[12pt,hidelinks]{article}
\usepackage[body={17.5cm, 22cm},right=2cm]{geometry}
\usepackage{color}
\usepackage{amssymb,graphicx}
\usepackage{amsmath}
\usepackage{epsfig}
\usepackage{lmodern}
\usepackage[normalem]{ulem}
\usepackage{bm}
\usepackage[bookmarks=true,pdfborder={0 0 0}]{hyperref} 
\usepackage[
    natbib=true,
    style=numeric-comp,
    sorting=none
]{biblatex}
\usepackage{float}
\allowdisplaybreaks

\definecolor{myorange}{RGB}{199,146,32}

\newcommand{\kdkJ}{\left.k_\delta/k_J\right|_{\rm MRE}}

\begin{document}
\setcounter{page}{0}
\thispagestyle{empty}

\parskip 3pt

\font\mini=cmr10 at 2pt

\begin{titlepage}
\noindent \makebox[15cm][l]{\footnotesize \hspace*{-.2cm} }{\footnotesize 
DESY-24-075}  \\  [-1mm]
~\vspace{2cm}
\begin{center}

{\LARGE \bf More Axion Stars from Strings}

\vspace{0.6cm}

{\large
Marco~Gorghetto$^a$,
Edward~Hardy$^b$,  
and Giovanni~Villadoro$^{c}$}
\\
\vspace{.6cm}
{\normalsize { \sl $^{a}$ Deutsches Elektronen-Synchrotron DESY, Notkestr. 85, 22607 Hamburg, Germany }}

\vspace{.3cm}
{\normalsize { \sl $^{b}$ Rudolf Peierls Centre for Theoretical Physics, University of Oxford, \\ Parks Road, Oxford OX1 3PU, UK}}

\vspace{.3cm}
{\normalsize { \sl $^{c}$ Abdus Salam International Centre for Theoretical Physics, \\
Strada Costiera 11, 34151, Trieste, Italy}}

\end{center}
\vspace{.8cm}
\begin{abstract}
We show that if dark matter consists of QCD axions in the post-inflationary scenario more than ten percent of it efficiently collapses into Bose stars at matter-radiation equality. Such a result is mostly independent of the present uncertainties on the axion mass. This large population of solitons, with asteroid masses and Earth-Moon distance sizes, might plausibly survive until today, with potentially interesting implications for phenomenology and experimental searches.

\end{abstract}

\end{titlepage}

{\fontsize{11}{10.8}
\tableofcontents
}

\section{Introduction} \label{sec:intro}

The QCD axion is a well-motivated extension to Standard Model of particle physics. In addition to being the most robust known solution to the strong CP problem \cite{Peccei:1977hh,Weinberg:1977ma,Wilczek:1977pj}, for values of the axion decay constant $f_a$ allowed by experiments it inevitably comprises a component of cold dark matter and it might make up the entirety of the observed dark matter abundance \cite{Preskill:1982cy,Abbott:1982af,Dine:1982ah}. Among the two broad classes of axion cosmological histories, the post-inflationary scenario \cite{Sikivie:1982qv,Vilenkin:1982ks,Davis:1986xc,Harari:1987ht,Linde:1990yj,Lyth:1992tw} has the distinguishing feature of being predictive and it also leads to interesting phenomenology.  Such predictions (the status of which we review in the next Section) and phenomenology could have important implications for the extensive experimental and observational program aiming to discover the QCD axion (see e.g.~\cite{Graham:2015ouw} and references therein) and therefore merit careful investigation.

Notably, QCD axion dark matter in the post-inflationary scenario is thought to automatically lead to dark matter substructure, i.e. gravitationally bound clumps of dark matter  \cite{Hogan:1988mp,Kolb:1993zz,Kolb:1994fi,Zurek:2006sy}. This substructure first forms in the early universe around the time of matter-radiation equality due to the collapse of isocurvature fluctuations. The resulting clumps of axions are expected to have densities comparable to the average dark matter density at the time of formation, roughly ${\rm eV}^4$, which is substantially larger than the present-day dark matter density in the vicinity of the Sun. Such substructure has been studied extensively both by analytical approaches and, most commonly, numerically  \cite{Fairbairn:2017dmf,Fairbairn:2017sil,Eggemeier:2019khm,Xiao:2021nkb,Ellis:2022grh,Eggemeier:2022hqa,Pierobon:2023ozb,Eggemeier:2024fzs}.

\begin{figure}[t]
    \centering
    \includegraphics[height=0.435\textwidth]{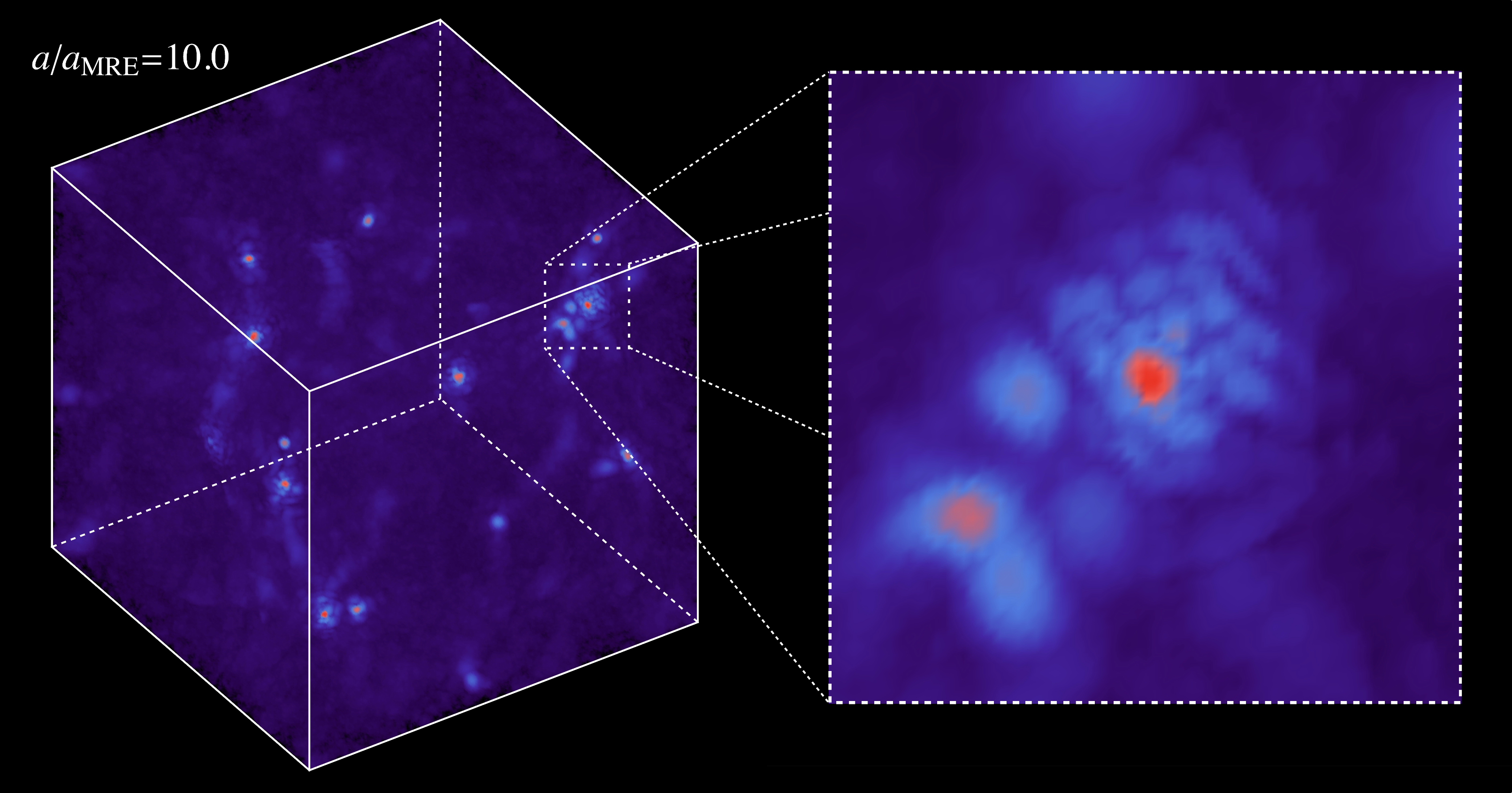}
     \includegraphics[height=0.4\textwidth]{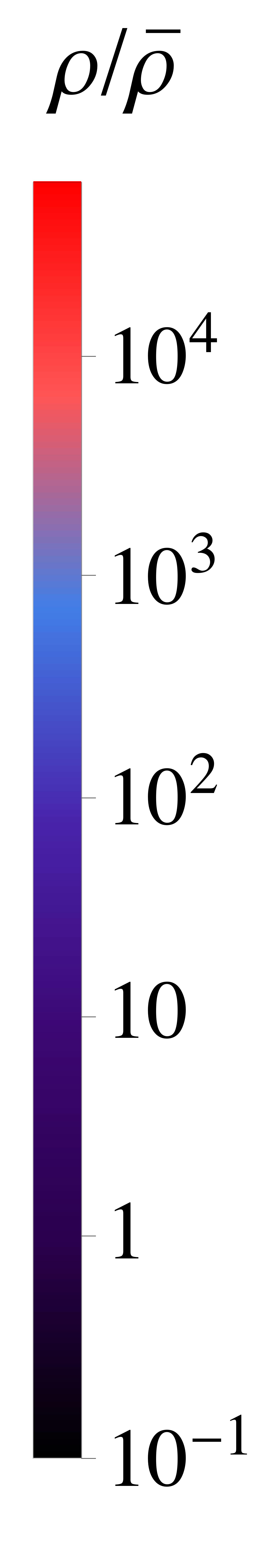}
    \caption{The axion energy density $\rho$, relative to its spatially averaged value $\bar{\rho}$, soon after matter-radiation equality in a numerical simulation. The initial conditions correspond to those expected for $f_a\simeq 2\times 10^{10}~{\rm GeV}$, with the peak of the density power spectrum $k_\delta$ such that $\kdkJ=2.4$ where $k_J$ is the quantum Jeans scale (results are similar for $f_a$ between $10^{10}~{\rm GeV}$ and $10^{11}~{\rm GeV}$ albeit with axion stars forming slightly later or earlier respectively). At the time shown, when the scale factor $a/a_{\rm MRE}=10$, approximately $15\%$ of the dark matter is in axion stars with densities greater than $10^4 \bar{\rho}$ (red objects). The zoomed-in view in the right panel shows the densest soliton in the simulation, which is in the process of merging with another one. Characteristic interference patterns from the wave-like nature of the dark matter can be seen.}
    \label{fig:pic}
\end{figure}

We will argue that the substructure takes a dramatically different form than previously thought, in particular that a large fraction of the clumps are actually \emph{axion stars}. These are solitons, gravitationally bound objects with size comparable to the de Broglie wavelength of their constituents, see e.g. refs.~\cite{Jetzer:1991jr,Niemeyer:2019aqm} for reviews. Axion stars have previously been believed to  mostly form  by relaxation via gravitational or self-interactions inside the dark matter clumps \cite{Tkachev:1991ka,Seidel:1993zk,Semikoz:1995rd}, therefore accounting for only a miniscule fraction of the total dark matter. We find instead that the axions in these solitons can comprise as much as one fifth of the total  dark matter, implying a much larger population of such objects.
An illustration of our results can be found in Figure~\ref{fig:pic}, which shows the axion energy distribution soon after matter-radiation equality in a numerical simulation.

The formation of axion stars at matter-radiation equality has been largely overlooked by previous numerical studies because they employed N-body simulations,
which are blind to the wave nature of axions.\footnote{Compatibly with our results, ref.~\cite{Eggemeier:2019jsu} also found that axion stars can form soon after matter-radiation equality, albeit in simulations with the axion mass $m\simeq 10^{-8}~{\rm eV}$, much smaller than the physical value  in the post-inflationary scenario.} 
The latter turns out to be important because of
a combination of different effects that 
shifts the spectrum of 
inhomogeneities towards smaller spatial scales  
than previously assumed (in particular, close to the axion de-Broglie wavelength). Some of these effects have being appreciated only recently thanks to modern simulations of the string network evolution~\cite{Gorghetto:2018myk,Gorghetto:2020qws,Buschmann:2021sdq,Saikawa:2024bta,Kim:2024wku}. One instead is new and is 
associated to the impact of self-interactions on non-relativistic axions as the axion potential grows after the decay
of topological defects, which turns out to be crucial 
for a reliable determination of the final axion spectrum and therefore the properties of the axion stars.

This paper is structured as follows: In Section~\ref{sec:review} we review the current status of understanding of the evolution of topological defects in the post-inflationary scenario. In Section~\ref{sec:new} we show that, between the time when the axion string-wall network is destroyed and the QCD crossover, self-interactions have a substantial effect on the axion energy spectrum. In Section~\ref{sec:mre} we consider the evolution of the axion field through matter-radiation equality and study the formation of axion stars. In Section~\ref{sec:stars} we discuss the possible dynamics of the axion stars long after matter-radiation equality, and in Section~\ref{sec:concl} we comment on directions for future work and possible observational and experimental implications. Supporting evidence for our results and details of our numerical simulations is provided in Appendices.

\section{Recap of the evolution of topological defects}\label{sec:review}
In the post-inflationary scenario (i.e. when the Peccei--Quinn (PQ)  symmetry is broken after inflation),
axion strings form and their dynamics dominate the  
field evolution until the axion potential becomes relevant, close to the QCD crossover temperature $T_c\simeq 155~{\rm MeV}$. The evolution of the string network at early times has been studied extensively
over the years mostly using numerical simulations \cite{Hagmann:1990mj,Battye:1993jv,Yamaguchi:1998gx,Yamaguchi:1998gx,Hagmann:2000ja,Hiramatsu:2010yu,Hiramatsu:2012gg,Kawasaki:2014sqa,Fleury:2015aca,Klaer:2017qhr,Klaer:2017ond,Gorghetto:2018myk,Kawasaki:2018bzv,Klaer:2019fxc,Gorghetto:2020qws,Buschmann:2021sdq,Saikawa:2024bta,Kim:2024wku}. The present understanding is that
soon after formation the string network is driven into an attractor of the evolution,
the scaling solution, independently of the initial conditions. 
The attactor is such that on average
the total string length per Hubble patch $\xi$ is fixed in terms of the ratio
of the Hubble parameter $H$ and the inverse string core size $m_r$ ($\xi\simeq 0.24\log(m_r/H)$).

To maintain the scaling solution the string network emits axion waves that populate
a gas of relativistic axions. At any given time the energy density of these free axions
is comparable to the energy density of the string network. The scaling regime ends when
the QCD axion potential starts to be relevant, when the temperature of the Universe
approaches the QCD scale. At this point domain walls form and, provided these are not stable, the string-wall network collapses into axion waves. After a transient in which non-linearities from the axion potential are important, the axions become non-relativistic and their comoving number density is 
conserved. 
The number of axions produced by strings and the string-domain wall network decay is strongly affected by the power 
spectrum of the emitted axions: depending on whether this is more UV or IR dominated
the final axion dark matter abundance from such processes could be negligible or enhanced with respect to the naive estimate of that from domain wall decay.  
A dedicated numerical study with 
high statistics was carried out in ref.~\cite{Gorghetto:2020qws}. It was found that the spectrum of axions emitted during the scaling regime
 is UV dominated at early times, but the spectral index evolves
logarithmically with time towards an IR dominated spectrum. 
Unfortunately the limited extent of the simulations did not allow 
this change in behavior to be confirmed, although the statistical precision of the data strongly disfavors a non-IR dominated spectrum at late times.  If an IR dominated spectrum is firmly established, 
the number of axions produced by strings is enhanced, pointing to values of $f_a\simeq 10^{10}$~GeV
or lower. Employing adaptive meshing techniques, a subsequent study was able to simulate the scaling solution for a longer time~\cite{Buschmann:2021sdq}. While the results are fully compatible with those in ref.~\cite{Gorghetto:2020qws}, the larger statistical errors 
of this study neither allowed the small-time evolution observed in \cite{Gorghetto:2020qws} to be resolved, nor an IR or UV dominated spectrum to be distinguished 
at the level required for a reliable extrapolation.  
Ref.~\cite{Saikawa:2024bta} recently performed large simulations obtaining high statistics with a wide range of initial conditions and carried out a thorough analysis of systematic uncertainties. The results confirm the evolution away from a UV dominated emission spectrum consistently with \cite{Gorghetto:2020qws}. However, increasing systematic uncertainties as spectral index $q=1$ is approached prevented a conclusive determination of the asymptotic emission spectrum. Meanwhile, ref.~\cite{Kim:2024wku} also studied the impact of different initial conditions in detail, finding strong evidence for the logarithmic evolution of $q$. 
To summarize, all high-statistics simulations to date agree on the presence of a
logarithmic evolution of the spectral index $q$, strongly disfavoring a UV dominated spectrum. The resulting preferred values for the axion decay constant are $f_a\in [1,6]\times 10^{10}$~GeV, with the two extrema corresponding to IR dominated and scale invariant
spectrum respectively. Lower values of $f_a$ are possible in the case the production from
the decay of domain walls dominates over that from strings or if the domain wall number $N_{\rm DW}>1$ \cite{Gorghetto:2020qws}.

While waiting for bigger simulations with higher statistics to improve the determination of $f_a$, we now discuss the evolution of the non-relativistic axions from the time when the strings and domain walls decay until the formation of the first gravitationally bound structures at around matter-radiation
equality (MRE), leaving $f_a$ as a free parameter pending a future definitive result. In particular, we will show how the smaller values of $f_a$, preferred by the most recent numerical simulations, affect the nature of the small-scale structures of QCD axion dark matter considerably.

\section{Evolution after topological defects decay}\label{sec:new}

Consider first the case of an axion-like particle (ALP) that has a temperature-independent mass $m$. Let us assume that at the time $t_\star$, defined by the condition $H_\star=H(t_\star)=m$, the ALP {energy density} spectrum is peaked at the momentum $k_{p\star}$. 
Of key importance to our work is the \emph{quantum} Jeans scale $k_J\equiv(16\pi G \rho m^2)^{1/4}$ (where $\rho$ is the axion energy density), which sets the  scale
below which  modes can collapse gravitationally, see Section~\ref{sec:mre} and refs.~\cite{Khlopov:1985fch,Hu:2000ke}. 
If we compare the value of the peak momentum at MRE, $k_{p}|_{{\rm MRE}}=k_{p\star} a_\star/a_{\rm MRE}$, where $a$ is the scale factor, with the quantum Jeans scale at that time, we find
\begin{equation} \label{eq:1}
    \left. \frac{k_{p}}{k_J}\right|_{\rm MRE}  = \frac{k_{p\star}a_\star/a_{\rm MRE}}{(16\pi G \rho_{\rm MRE} m^2)^{1/4}} \simeq \frac{k_{p\star}}{H_\star} \,,
\end{equation}
where we used that the axion energy density at MRE is $\rho_{\rm MRE}\simeq \rho_{\rm MRE}^{\rm (SM)}\simeq \rho_\star^{({\rm SM})} (a_\star/a_{\rm MRE})^{4}$ and 
$\rho_\star^{({\rm SM})}=3H_\star^2/(8\pi G)$ (we omit an order-one factor on the right hand side of eq.~\eqref{eq:1}, but include this in our subsequent numerical results). This means that if the spectrum of ALPs is originally peaked at $H_\star$, at MRE
its peak is close to the 
Jeans scale. In other words, the 
dark matter fluctuations 
that first gravitationally collapse (at around MRE) have a size comparable to the typical de Broglie
wavelength of the ALPs, i.e. they are Bose condensates. In this case a fraction of dark matter would form a large number of Bose stars already around MRE. This fact was noticed before in the context of dark photon dark matter in ref.~\cite{Gorghetto:2022sue},
but as we just saw it can happen anytime a spectrum of bosonic dark matter is produced peaked at the Hubble scale at the time it becomes non-relativistic. 
In reality, for an ALP in the post-inflationary scenario 
$k_{p\star}$ is expected to be one order of magnitude or more larger than $H_\star$ and the production of Bose stars will not be efficient~\cite{Gorghetto:2022ikz}, as we are going to see later.

For the case of the QCD axion there are several differences. The main one is 
that the axion mass is temperature-dependent and continues to {grow} even after $t_\star$. 
The relation between $H_\star$ and the zero-temperature mass appearing in $k_J$ at MRE will therefore differ. 
Repeating the same steps as before but keeping track of the difference between the late-time mass $m$ and $m_\star=m(t_\star)$ we have
\begin{equation}
    \left. \frac{k_{p}}{k_J}\right|_{\rm MRE}  = \frac{k_{p\star}a_\star/a_{\rm MRE}}{(16\pi G \rho_{\rm MRE} m^2)^{1/4}}\simeq \frac{k_{p\star}}{H_\star}\left(\frac{m_\star}{m}\right)^{1/2} \,.
\end{equation}
Therefore, given that $m_\star/m\simeq (T_c/T_\star)^4\sim (f_a/M_p)^{2/3}$ \cite{Gross:1980br,Borsanyi:2016ksw},
\begin{equation}\label{eq:naive}
    \left. \frac{k_{p}}{k_J}\right|_{\rm MRE}  \simeq  \left(\frac{f_a}{M_p}\right)^{1/3} \frac{k_{p\star}}{H_\star} \sim 10^{-3}\, \frac{k_{p\star}}{H_\star} \,.
\end{equation}
This would imply that for the QCD axion the spectrum is peaked at length scales much larger than the Jeans scale and the gravitationally bound structures that would form 
at MRE more closely resemble virialized halos of particles, {\it miniclusters}, than solitonic bound states {\it axion stars}. 

We are going to challenge this standard lore and argue that the naive estimate in eq.~\eqref{eq:naive} is not correct. This is because 
the time-dependence of the axion potential affects the evolution of axions non-trivially even after they become non-relativistic, and non-linearities, although 
too small to affect the conservation of number density, still play a crucial role in reshaping the axion spectrum.\footnote{\label{ft:oscillon}The destruction of the string-wall network will produce some long-lived oscillons called `axitons' \cite{Kolb:1993hw,Vaquero:2018tib} (quasi-stable configurations with inverse size $m$ in which the axion field $\phi\sim f_a$), which can lead to additional inhomogeneities on small scales when they decay. {Axitons are also the result of the self-interaction, but they act on smaller scales} and are therefore distinct from the processes that we focus on in which the axion field remains in the non-relativistic regime and with amplitude much smaller than $f_a$.} 

To understand why this is the case, it is useful to track the importance of each
term of the Hamiltonian as a function of time. At $T=T_\star$ the mass, self-interaction and gradient energy densities are roughly similar. For ALPs with a constant mass, in the
non-relativistic regime (i.e. after the mass term starts dominating the energy density) 
the mass term redshifts as $m^2 \phi^2 \propto T^3$ (from the conservation of the comoving number density $n$ of ALPs, $n\sim m\phi^2\propto s \propto T^3$, where $\phi$ is the axion field), the gradient term redshifts as $(\nabla \phi)^2\propto T^5$ and the quartic self-interaction as $\lambda \phi^4\propto T^6$
(higher order non-linearities will redshift even faster; here $\lambda\equiv -V''''(0)$). Therefore, after the field becomes
non-relativistic the hierarchy $m^2\phi^2 \gg (\nabla \phi)^2\gg \lambda \phi^4$ develops.
The first of these inequalities signals that the ALPs become more and more non-relativistic, while the second shows that
the self-interactions become less and less relevant.  
In principle, in this regime  
the self-interactions affect the spectrum by transferring momentum into the UV (the usual UV catastrophe of
classical field theory) on timescales $\tau_{\rm therm}=64
m^5 k_p^2
/\left(\lambda^2 \bar{\rho}^2\right)$ \cite{Levkov:2018kau}, where $\bar{\rho}$ is the average energy density.
However, the thermalization process 
rapidly freezes out due to the Hubble expansion because 
\begin{equation}\label{eq:scatterALP}
\tau_{\rm therm}H \ \simeq \ 64\left( \frac{ m f_a^2}{n (T_\star/T)^3} \right)^2\left(\frac{T_\star}{T}\right)^2 \left(\frac{k_{p\star}}{H_\star}\right)^2\gg 1~,
\end{equation}
for $T\ll T_\star$.

For the QCD axion the steep time-dependence of the potential changes the relative importance of
the various terms. As long as $T>T_c$ the axion mass increases as the Universe cools
as $m\propto T^{-4}$. Similarly, the quartic coupling increases as $\lambda\sim (m/f_a)^2\propto T^{-8}$.
After the axions become non-relativistic the number density is still covariantly conserved,
therefore $m\phi^2\propto T^3$, which now implies that $\phi^2\propto T^7$. We therefore have
that the mass term now redshifts as $m^2\phi^2\propto T^{-1}$, the quartic as 
$\lambda \phi^4 \propto T^6$ while the gradient $(\nabla \phi)^2\propto T^9$ (this time-dependence is illustrated in  Figure~\ref{fig:energy_evo} of Appendix~\ref{app:self_analytic}). Consequently the 
hierarchies are different: $m^2\phi^2 \gg \lambda \phi^4 \gg (\nabla \phi)^2$. The field
is still non-relativistic, which means that the comoving number density is conserved, 
and higher non-linearities are successively smaller. However, the kinetic energy of the axion
gas is small compared to the  self-interaction energy. When this happens the kinetic pressure is
not able to balance the self-attraction of the axions, which start clumping, and they accelerate until their kinetic energy becomes comparable to the self-interaction energy, i.e. they virialize.

As we subsequently confirm with numerical simulations, in the regime $m^2\phi^2 \gg \lambda \phi^4 \gg (\nabla \phi)^2$ energy is moved to the UV on timescales of order \cite{Sikivie:2009qn} 
\begin{equation}\label{eq:tnl}
\tau_{v}= 8m/(\lambda \phi^2) ~.
\end{equation}
For $T_c<T<T_\star$  
\begin{equation} \label{eq:tauvH}
\tau_vH\simeq \left(\frac{m_\star f_a^2}{n (T_\star/T)^3}\right) \left(\frac{T_\star}{T}\right) \simeq  0.5 \left(\frac{f_a}{10^{10}\,{\rm GeV}}\right)\left(\frac{T_c}{T}\right)~,    
\end{equation}
where we used that the axion comprises the full dark matter abundance in the second equality.
Consequently, $\tau_v$ is fast on cosmological timescales  for some range of $T\lesssim T_\star$  
provided $f_a\lesssim 10^{11}\,{\rm GeV}$. As a result, the peak momentum of the spectrum is driven to larger values, close to the `critical' virialized momentum 
$k_v=\sqrt{\lambda  \phi^2 
}$, such that $(\nabla \phi)^2\sim \lambda \phi^4$. 
Notably $\tau_{\rm therm}\gg \tau_v$ for $k_p\gg k_v$ (eq.~\eqref{eq:scatterALP} with $m\rightarrow m_\star$ applies also for the QCD axion in this regime), so $k_v$ acts as an approximate attractor. These dynamics, with $k_p$ tracking the (time-dependent) $k_v$, last until the QCD axion potential stops growing around $T\sim T_c$. Soon after that the normal ALP hierarchy among the terms in the Hamiltonian is restored and the self-interactions freeze out. 

We therefore assume that for the QCD axion 
$k_p\sim k_v$ is maintained until $T\sim T_c$ and only afterwards $k_p$ redshifts
freely. At $T=T_c$ the peak momentum $k_p\sim m\phi/f_a\sim\sqrt{\rho(T_c)}/f_a$.
Using that $\rho(T_c)=\rho_{\rm MRE}(a_{\rm MRE}/a_c)^3$, we then have  
$k_{p}|_{\rm MRE}=\sqrt{\rho_{\rm MRE}}(a_{\rm MRE}/a_c)^{1/2}/f_a$. As a result, we estimate
\begin{equation} \label{eq:estimate}
\left. \frac{k_{p}}{k_J}\right|_{\rm MRE}  \sim  
\frac{\sqrt{\rho_{\rm MRE}}(a_{\rm MRE}/a_c)^{1/2}/f_a}{(16\pi G \rho_{\rm MRE} m^2)^{1/4}} 
\sim
\left(\frac{M_p}{f_a} \frac{T_{\rm MRE} }{T_c} \right)^{1/2}\sim 
\left(\frac{10^{10}~{\rm GeV}}{f_a}\right)^{1/2} ~,
\end{equation}
where we used the approximate parametric relations $\rho_{\rm MRE}\sim T_{\rm MRE}^4$ and 
$m\sim T_c^2/f_a$. Consequently, for values of $f_a$ in the range that might lead to QCD axion dark matter, the peak of the axion energy density spectrum is also close to the quantum Jeans scale at MRE!

The rough estimate in eq.~\eqref{eq:estimate} captures the main features of the dynamics,  but neglects several effects that partially limit its applicability. 
For the larger values of $f_a$ (i.e. around $10^{11}$~GeV) $\tau_vH\gtrsim 1$ at some intermediate $T_c\lesssim T\lesssim T_\star$ so that self-interactions freeze-out before $T=T_c$. Moreover, given that the axion kinetic energy is initially larger than the quartic term (because, as observed from simulations, $k_p/H_\star={\cal O}(10)$ at $t_\star$) the self-interaction term might not have caught up to the gradient term prior such times. 
In this case $k_p/k_J|_{\rm MRE}$ remains close its initial value set by the string-wall decay, which for such $f_a$ is likely to be ${\cal O}(10^{-1})$ depending on the shape of the spectrum emitted by strings.

The situation is also different at smaller $f_a$ (in particular, $f_a\lesssim 5\times 10^{10}$~GeV). For these values, 
the axion abundance from strings is so large  
that non-linearities delay the onset of the non-relativistic regime until a later time $t_\ell$ (when $T=T_\ell< T_\star$) and shift $k_p$ towards the UV, to the momentum that matches the axion mass at $t_\ell$~\cite{Gorghetto:2020qws}. The physics at $T_\ell$
is dominated by relativistic non-linear dynamics and it is at this time that (topologically trivial) domain walls will decay into a gas of axions that rapidly becomes non-relativistic.
At the start of the non-relativistic regime the quartic term is only slightly smaller than the gradient one, so it again starts dominating before $T_c$, entering a second
phase of non-linear dynamics, this time in the non-relativistic regime. The combination of the first UV shift during the relativistic non-linear evolution at $t_\ell$ and the second non-relativistic one still results in $\left.k_p/k_J\right|_{\rm MRE}\simeq 1$.

\subsection{Numerical evolution before matter-radiation equality} \label{sec:simpre}

To explore the dynamics in the regime  $m^2\phi^2 \gg \lambda \phi^4 \gg (\nabla \phi)^2$, we numerically evolve realizations of the axion field on a discrete lattice. We do this both in flat space-time (with constant axion mass, quartic coupling constant, and scale factor), and cosmological simulations during radiation domination, starting from when the axion field is first non-relativistic, not much after $T=T_\star$,  to times when $T\lesssim T_c$ and the self-interactions have frozen out. 
We define the non-relativistic field $\psi$ by
\begin{equation}\label{eq:phi_t}
\phi= \frac{1}{\sqrt{2 m_0m a^3}}\left(\psi e^{-i\int^t m(t')dt'}+{\rm c.c.}\right) \, ,
\end{equation}
where $m_0$ is the zero-temperature axion mass and $a\propto t^{1/2}$ is the scale factor. In the limit $\dot{\psi}\ll m\psi$, $\ddot{\psi}\ll m\dot{\psi}$, $H\ll m$,  
which are satisfied soon after $T$ drops below $T_\star$, the axion's equation of motion becomes
\begin{equation}\label{eq:s-tdep}
	\left(i\partial_t+\frac{\nabla^2}{2m}-m\Phi + \frac{\lambda|\psi|^2}{8a^3m_0m^2}\right)\psi=0 \, ,
\end{equation}
where spatial derivatives are with respect to physical distances, and we expand the axion's potential $V=\frac{1}{2}m^2\phi^2 - \frac{1}{4!}\lambda \phi^4+\ldots$ such that $\lambda> 0$ for an attractive self-interaction, as for the QCD axion.  
At $T\sim T_c$ the gravitational potential $\Phi$ sourced by the axion field  is negligible.
We fix the initial field to have a Gaussian distribution,\footnote{In fact, rather than represent a pure gas of uncorrelated waves, the axion field is expected to have non-Gaussian features associated to the 
to the non-linear transient at $t_\ell$ and the decay of the string-wall network \cite{Vaquero:2018tib}, but we do not expect our quantitative conclusions to be substantially affected.} with 
power spectrum $\mathcal{P}_\phi$ peaked at $k_p$ and with $\mathcal{P}_{\dot{\phi}}(k)=m^2 \mathcal{P}_{\phi}(k)$, where for generic field $X(t,\vec{x})$ we define
\begin{equation}\label{eq:Pdelta}
\langle X^*(t,\vec{k})X(t,\vec{k'})  \rangle = \left(2\pi\right)^3 \delta^{(3)}\left(\vec{k}-\vec{k}' \right) \frac{2\pi^2}{k^3} \mathcal{P}_X(t,k) ~,
\end{equation}
with $X(t,\vec{k})$ the Fourier transform of $X$. We have $m^2 \mathcal{P}_{\phi}(k)=\partial \rho/\partial \log k$ where $\rho$ is the total axion energy density, which is dominated by the mass energy at these times. {Each momentum mode is generated with a random phase. The box has typical size $2\pi/k_p$, and thus contains only a fraction of the Hubble volume.}  
As we discuss subsequently, the specific form of the initial spectrum at $k\ll k_p$ and $k\gg k_p$ is unimportant for our results; we take $\mathcal{P}_{\phi}(k) \propto k^3$ for $k\ll k_p$ and $\propto k^{-1}$ for $k\gg k_p$.

Simulations in flat space-time confirm that while $k_p \lesssim k_v$ the energy spectrum  indeed evolves towards the UV on timescales of order $\tau_v$ given by eq.~\eqref{eq:tnl}.  This is in fact the only time scale of the equations of motion, eq.~\eqref{eq:s-tdep}, if the gradient and the gravitational potential terms are negligible.
Once the peak of the energy spectrum reaches $k_p\gtrsim k_v$ the gradient term becomes relevant and the evolution slows, consistent with the expression for $\tau_{\rm therm}$, which is valid in this regime. Meanwhile, 
cosmological simulations show that the effects are as anticipated: for $f_a\lesssim 10^{11}~{\rm GeV}$,  $k_p$ is driven close to $k_v$ while $T \gtrsim T_c$ and once $T\lesssim T_c$ the comoving spectrum freezes. The resulting attractor-like behaviour is imperfect because $k_v$ is time-dependent and even once $k_p\simeq k_v$ there is a slow drift of energy into the UV on timescales of order $\tau_{\rm therm}$. The self-interactions can alter the shape of the spectrum around its peak from its initial form by an order-one amount. We also find that the final value of $k_p$ is insensitive to the detailed shape of the initial spectrum at $k\ll k_p$ and $k\gg k_p$. Details and further analysis can be found in Appendix~\ref{app:simearly}.

\begin{figure}[t]    
    \centering
    \includegraphics[width=0.75\textwidth]{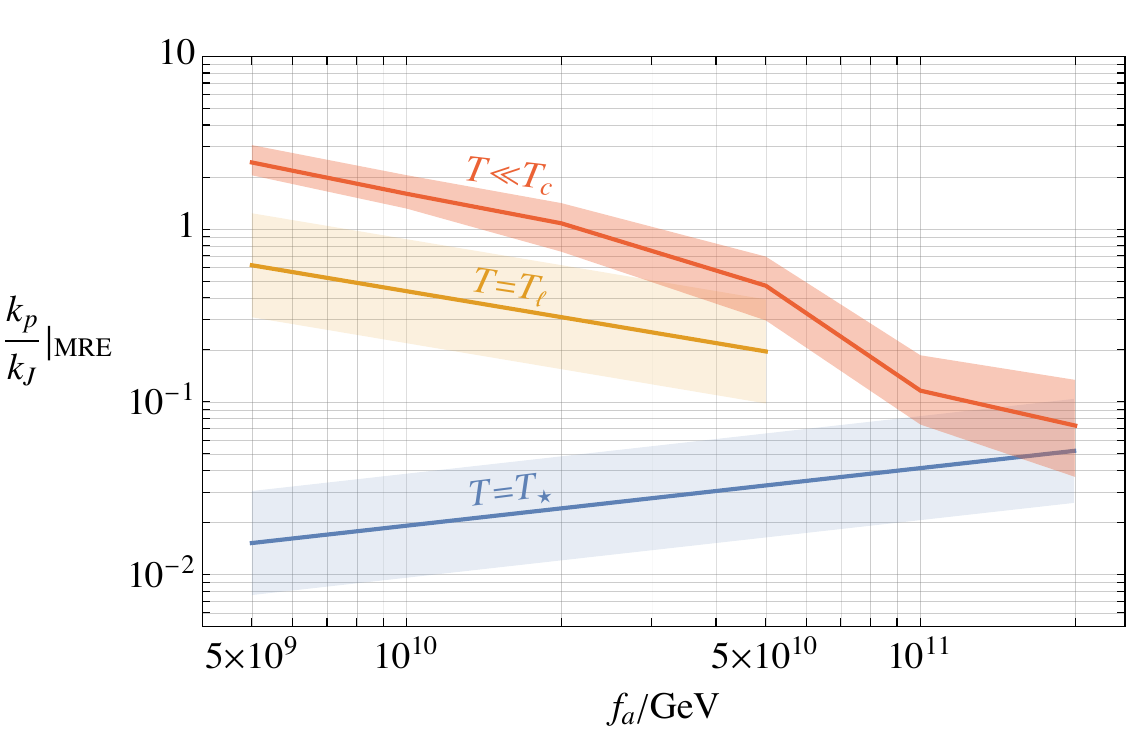}
    \caption{\label{fig:kpfakJ} The peak $k_p$ of the axion kinetic energy spectrum $\partial \rho/\partial \log k$ relative to the quantum Jeans scale at matter-radiation equality, $k_{J}|_{\rm MRE}$, as a function of $f_a$
    . {Different lines correspond to the peak momentum at different times in the cosmological history, redshifted freely to MRE.} Immediately after the decay of the string and domain wall network $k_p(T_\star)\sim 10 H_\star$ is expected (blue, labelled ``$T=T_\star$"). For $f_a\lesssim 5\times 10^{10}~{\rm GeV}$ there is a  non-linear transient soon after $T=T_\star$ as the axion field becomes non-relativistic, which increases $k_p$ to the orange line, ``$T=T_{\ell}$". Finally, at times around the QCD crossover the axion self-interactions increase $k_p$ to the red line ``$T\ll T_c$".}
\end{figure}

The values of $k_p$ at the different epochs are summarized in Figure~\ref{fig:kpfakJ} as a function of $f_a$. 
Although poorly known because a reliable extrapolation to large $\log(m_r/H)$ is lacking, we assume that the decay of the string network leads to $k_p(T_\star)\sim 10 H_\star$ and estimate the  uncertainty as $k_p(T_\star)/H_\star \in [5,20]$.\footnote{In more detail, numerical simulations at earlier times find $k_p(T_\star)\sim 10 H_\star$, which might be expected to increase proportionally to $\sqrt{\xi}$ during
the scaling regime \cite{Vaquero:2018tib}. However the non-linear dynamics at $t_\ell$ shift the spectrum to a larger value making the sensitivity on the original position of $k_p$ only logarithmic \cite{Gorghetto:2020qws}. The subsequent non-relativistic non-linear evolution until
$t_c$ discussed here further washes out the uncertainty related to $k_p(T_\star)$.}  
For $f_a\lesssim 5\times 10^{10}~{\rm GeV}$ the non-linear evolution of the axion field as it becomes non-relativistic at $T=T_{\ell}$ increases $k_p$. 
The value of $k_p$ immediately after this transient is approximately independent of $k_p(T_\star)$, however we assume a factor of $2$ uncertainty in its determination from simulation results presented in ref.~\cite{Gorghetto:2018myk}.\footnote{These results apply both if the instantaneous emission spectrum from strings is IR dominated at large $\log(m_r/H)$ or scale invariant, because in both cases the total axion spectrum at $T=T_\star$ has the same $1/k$ form up to logarithmic corrections. The amplitude of this spectrum is set by the assumption that, for a given $f_a$, the axion comprises the full dark matter abundance (e.g. by assuming different values of $\xi_\star$ or $\log(m_r/H_\star)$). 
For $f_a\gtrsim 5\times10^{10}~{\rm GeV}$ such a transient is absent. 
} 
Predictions for $k_p$ at $T\ll T_c$, after the axion self-interactions freeze out, are obtained from numerical simulations through the time when $T=T_c$. For each $f_a$ the central value of $k_p$ at $T\ll T_c$ is obtained from initial conditions with the $k_p$ expected from the previous evolution  (see Figure~\ref{fig:kikf} in Appendix~\ref{app:simearly}). The upper and lower uncertainties on the late-time $k_p$ correspond to initial conditions at the edges of the allowed ranges of $k_p$ after the prior evolution, with an additional $\pm 10\%$ error added to reflect the uncertainty in extracting the position of the final value of $k_p$ and systematic errors. 
For $f_a\lesssim 5\times 10^{10}~{\rm GeV}$ the self-interactions increase $k_p$ substantially  and the attractor-like behaviour reduces the relative uncertainty on $k_p$. As $f_a$ decreases below $10^{10}~{\rm GeV}$, the relation $k_p/k_J|_{\rm MRE}\propto f_a^{-1/2}$ from the estimate in eq.~\eqref{eq:estimate} becomes more and more accurate. Meanwhile for  $f_a\gtrsim 10^{11}~{\rm GeV}$ the axion self-interactions change $k_p$ only by an order-one factor, and the final $k_p$ has a stronger dependence on the initial conditions. 
We conclude that for $f_a\lesssim 5\times 10^{10}~{\rm GeV}$, as favoured by simulations, $k_p/k_{J}|_{\rm MRE} \gtrsim 0.5$. Even for $f_a\sim  10^{11}~{\rm GeV}$ at MRE $k_p$ is no more than an order of magnitude smaller than $k_{J}$.

\section{Evolution  around matter-radiation equality} \label{sec:mre}

After the  axion self-interactions freeze out at $T\simeq T_c$, and while the Universe is deep in radiation domination, the axion field evolves freely and its comoving energy spectrum stays fixed.\footnote{{Contrary to the adiabatic perturbations, our Gaussian isocurvature fluctuations are not affected by free-streaming and the corresponding $\mathcal{P}_
\delta$ is not expected to change in comoving coordinates from $T
\simeq T_c$ to $T\simeq T_{\rm MRE}$, see Ref.~\cite{Amin:2022nlh,Liu:2024pjg}.}} Owing to the non-zero momentum of the axion waves, the axion energy density inevitably has fluctuations (relative to the Standard Model radiation bath these are isothermal perturbations). The fluctuations are characterised by the density power spectrum $\mathcal{P}_\delta$, where $\delta(t,\vec{x})\equiv (\rho(t,\vec{x})- \bar{\rho}(t) )/\bar{\rho}(t) $ is the overdensity field. 
Assuming that the non-relativistic axion field is Gaussian, $\mathcal{P}_\delta$ can be straightforwardly expressed in terms of $\partial\rho/\partial\log k$ (see e.g. ref.~\cite{Enander:2017ogx}). We define $k_\delta$ to be the momentum at which $\mathcal{P}_\delta$ is maximized. For $\partial\rho/\partial\log k$ with the typical shape that emerges from the string decay and non-linear evolution, 
$k_\delta\simeq 2 k_p$ and $\mathcal{P}_\delta(k_\delta)\sim\mathcal{O}(1)$ indicating that there are order-one overdensities on spatial scales $\sim k_\delta^{-1}$. We note that $\mathcal{P}_\delta \propto k^3$ for $k\ll k_\delta$ provided  $\partial \rho/\partial\log k\propto k^\beta$ with $\beta\geq3/2$  for $k\ll k_p$, and $\mathcal{P}_\delta \propto \mathcal{P}_a$ for $k\gg k_\delta$. 
The results of numerical simulations seem compatible with
a $\mathcal{P}_\delta \propto k^3$ behavior in the IR part 
although the limited range of momenta available did not allow us to determine precisely such power. However, as will become clear later, 
the evolution of the system at around MRE is mostly determined by 
the position of the peak $k_\delta$ and much less by the precise values of
the power indices of the spectrum away from the peak.

The subsequent evolution of an overdensity depends on its size relative to the quantum Jeans scale $k_J$: overdensities on spatial scales much smaller than $k_J^{-1}$ are strongly affected by wave-effects, in particular quantum pressure, and those on spatial scales much larger than $k_J^{-1}$ are unaffected (this is most easily seen after a Madelung transformation of the equations of motion to a fluid description, as we review in Appendix~\ref{app:stars}) \cite{Hu:2000ke,Hui:2021tkt}. What is relevant for the collapse of a particular overdensity is the local value of $k_J$, but this is proportional to $(\rho(\vec{x})/\bar{\rho})^{1/4}$ so for the order-one to -ten overdensities typical of the initial axion field $k_J(\rho(\vec{x}))$ is  within a factor of a few of $k_J(\bar{\rho})$. Notably, the comoving quantum Jeans scale associated to the mean dark matter density $k_J(\bar{\rho}) a \propto a^{1/4}$ increases with the scale factor but only slowly. 

An overdensity that is unaffected by quantum pressure and has initial magnitude $\delta\gtrsim 1$ remains approximately frozen in comoving coordinates during radiation domination until $a/a_{\rm MRE}\simeq 1/\delta$   
when it undergoes gravitational collapse. Meanwhile fluctuations of initial $\delta\ll 1$ grow as $\delta(a)\propto (1+(3a)/(2a_{\rm MRE}))$ \cite{Meszaros} and collapse once they reach $\delta\simeq 1$.  The result of collapse is a minicluster supported by angular momentum 
that is expected to have a density of approximately $10^2 \delta^3(1+\delta)\bar{\rho}_{\rm MRE}$.  
N-body simulations suggest that the density profiles in the centers of such objects have a power law or Navarro-Frenk-White (NFW)~\cite{Navarro:1995iw} form \cite{Gosenca:2017ybi,Eggemeier:2019khm}.  
Conversely, overdensities that are affected by quantum pressure  oscillate rather than growing or collapsing. As a result, fluctuations on comoving spatial scales much smaller than $k_J^{-1}(\bar{\rho}_{\rm MRE})/a_{\rm MRE}$ do not collapse prior to structure formation on larger scales.

From Figure~\ref{fig:kpfakJ}, we see that for the QCD axion $k_p$, and consequently also $k_\delta \simeq 2k_p$, is within a factor of a few of $k_J(\bar{\rho})$ at MRE.  In this intermediate regime, quantum pressure is relevant on scales close to the size of the order-one overdensities and is therefore expected to play a role in the bound objects that form but not prevent collapse entirely.\footnote{In more detail, at the would-be time of collapse in the absence of quantum pressure, $a/a_{\rm MRE}=1/\delta$, the comoving quantum Jeans scale locally to an overdensity $\delta(\vec{x})\gtrsim 1$ is given by $k_J((1+\delta)\bar{\rho})= k_J(\bar{\rho})|_{\rm MRE}$. Hence, an overdensity on comoving scale $k_\delta a$ such that $(k_\delta/k_J(\bar{\rho}))|_{\rm MRE}<1$ is expected to collapse at $1/\delta$ as in the absence of quantum pressure. Meanwhile, for $(k_\delta/k_J(\bar{\rho}))|_{\rm MRE}> 1$ collapse occurs at $a/a_{\rm MRE}\simeq(k_\delta/k_{J,{\rm MRE}})^4/\delta$.}  
There are indeed solutions of the axion equations of motion and the Poisson equation consisting of gravitationally bound objects, \emph{axion stars}, that are supported by quantum pressure \cite{Ruffini:1969qy,PhysRevA.39.4207} (see also e.g. \cite{Chavanis_2011} for a recent discussion). In particular, we consider axion stars that are bound by gravitational interactions (as opposed to self-interactions) and in which the axions are non-relativistic. 
 The density profile of such an axion star takes the universal form
\begin{equation} \label{eq:axstarprof}
\rho(r)= \rho_s \chi\left(r/\lambda_J(\rho_s)\right)~,
\end{equation}
where $\chi(0)=1$, so $\rho_s$ is the central density, and $\lambda_J(\rho_s) \equiv 2\pi/k_J(\rho_s)$. The function $\chi(x)$ is close to constant for $x \ll 1$ and decays exponentially for $x\gg 1$. The de Broglie wavelength in the center of an axion star is of order $\lambda_J(\rho_s)$ and roughly 98\% of a star's mass is within this distance of its center. The mass of a star $M_s$ and the radius $R_{0.1}$ at which the density is a factor ten smaller than at the center (within which  approximately three quarters of the total mass is contained) satisfy 
\begin{align}
M_sR_{0.1} &\simeq \frac{5.2}{Gm^2} \implies R_{0.1}\simeq 
4.2\times 10^{6}~ {\rm km} \left( \frac{f_a}{10^{10}~{\rm GeV}} \right)^{2}  \left( \frac{10^{-19}M_\odot}{M_s} \right)~, 
\label{eq:starR}\\
\rho_s&\simeq 0.0044\, G^3M_s^4m^6  \simeq 
7.1\times 10^{-3}~{\rm eV}^4 \left( \frac{10^{10}~{\rm GeV}}{f_a} \right)^{6}  \left( \frac{M_s}{10^{-19}M_\odot} \right)^4 ~, 
\label{eq:starrho}
\end{align}
where we specialize to a QCD axion in relating $m$ and $f_a$. 
An axion star is the lowest energy configuration of a system with fixed particle number, and for an overdensity of size $k_J^{-1}$ the timescale for an axion star to form coincides with the gravitational in-fall time.

We therefore expect that for a QCD axion a substantial fraction of the bound objects that form from the collapse of the $O(1)$ density perturbations at around MRE are axion stars. These are likely to be surrounded by a ``fuzzy halo'' of axions that is partly supported by angular momentum.  
The central density of an axion star will be roughly given by the local axion density at the time when it forms, i.e. $\rho(t_{\rm coll},\vec{x})= (1+\delta)\bar{\rho}_{\rm MRE} \left(a_{\rm MRE}/a_{\rm coll}\right)^3$ where  ``${\rm coll}$'' denotes quantities at the time of collapse. Meanwhile, the mass of an axion star $M_s$ is expected to be an order-one fraction of the total mass in the initial fluctuation. As a result, the order-one overdensities are expected to lead to axion stars of mass 
\begin{equation}
\begin{aligned} \label{eq:Mexp}
M_s&\simeq (1+\delta) \bar{\rho}(t_{\rm coll}) \left(2\pi/k_{\delta}(t_{\rm coll})\right)^3 \\ &\simeq (1+\delta) \left. \left(\frac{k_{J}}{k_\delta}\right)^3\right|_{\rm MRE}  M_{J,{\rm MRE}} ~,
\end{aligned}
\end{equation}
 where we define $M_{J,{\rm MRE}}= \bar{\rho}_{\rm MRE}\left(2\pi/k_{J,{\rm MRE}}\right)^3 \simeq 2.2\times 10^{-19}M_\odot \left(f_a/10^{10}~{\rm GeV}\right)^{3/2}$. The masses predicted by eq.~\eqref{eq:Mexp} are self-consistently such that the axion quartic coupling is negligible in the axion stars.

\subsection{Numerical simulations around matter-radiation equality} \label{sec:simpost}

To determine whether axion stars do indeed form, we numerically solve the equations of motion of the non-relativistic axion field through MRE with initial conditions with different $\left. k_\delta/k_{J}\right|_{\rm MRE} \in[0.3,5]$ corresponding to the plausible range of $k_p$ identified in Section~\ref{sec:new}. 
The set-up of these simulations is similar to those described in Section~\ref{sec:simpre},  
but with the gravitational potential $\Phi$ included.  
We start the evolution at $a/a_{\rm MRE}=0.01$, when the spectrum of density fluctuations is frozen, with an initial axion field that is Gaussian with energy spectrum close to the expectation from the earlier evolution, in particular given by eq.~\eqref{eq:shape} in Appendix~\ref{app:simsetup} with $s=4$. The simulation results are only reliable while 1) the (increasing) physical lattice spacing is sufficient to resolve the cores of collapsed objects and 2) the density fluctuations on length scales comparable to the box remain perturbative, i.e. $\mathcal{P}_\delta\left(2\pi/L\right)\lesssim 1$ where $L$ the box size. The maximum scale factor compatible with the preceding, competing, requirements given our available computing resources depends on the initial $k_\delta$.

For all initial $k_{\delta}$ tested, density perturbations collapse into gravitationally bound objects around the time of MRE, characterised by their central density decoupling from Hubble expansion and instead remaining approximately constant.  At the final simulation times these objects are mostly well-separated, although  the beginnings of a ``cosmic web'' of overdense filaments is visible in Figure~\ref{fig:pic}. Gravitationally bound objects form later for larger $k_\delta$, which is consistent with quantum pressure delaying collapse (see Appendix~\ref{app:sim_mre}). We confirm that the axion quartic self-interactions play no role in the dynamics by carrying out simulations starting from identical initial conditions with  $\lambda=0$ and $\lambda$ set to its physical value, which lead to final field configurations that differ by much less than $ 1\%$.

We classify gravitationally bound objects as axion stars or miniclusters based on their density profiles. First, we define an object as collapsed if its central density $\rho_s>50 \bar{\rho}(t)$, which in practice captures  all objects that have approximately constant density. We then identify an object as being an axion star if its spherically averaged density profile matches the predicted form, eq.~\eqref{eq:axstarprof}, to within a factor of $2$ at both $\lambda_J(\rho_s)/4$ and $\lambda_J(\rho_s)/2$; at these points the predicted axion star $\rho(r)$ is  approximately 60\% and 16\% of $\rho_s$, respectively. We use these generous identification criteria to account for the stars not forming in the ground state immediately (as we discuss at the end of the Section, demanding stronger criteria does not change our conclusions substantially).  The radial derivative of the quantum pressure $\partial_r\Phi_Q$ matches that of the gravitational potential $\partial_r \Phi$ to within a factor of $2$ at both $\lambda_J(\rho_s)/2$ and $\lambda_J(\rho_s)/4$ in roughly half of the objects identified as stars (with slightly larger deviations in the remainder, which are typically recently formed), confirming that quantum pressure indeed plays a role in supporting the objects. 
The regions identified as stars are close to spherically symmetric, with the projections of the density field onto spherical harmonics $y_{l}^m$ satisfying $\int d\Omega~ y_{l}^m(\theta,\phi)\rho(r,\theta,\phi) \lesssim  \rho(r)/10$ for $r\lesssim \lambda_J(\rho_s)$ with $l\geq 1$.

Based on the preceding definition, more than $75\%$ of identified objects contain a central axion star for all initial $k_\delta$. As expected, the axion stars are surrounded by a halo in which the de Broglie wavelength is comparable to the typical length-scale but angular momentum is relevant. Such fuzzy halos are evident from the density profiles deviating from the axion star prediction and instead taking a power law form $\rho(r)\propto r^{-n}$ with $n\simeq 2.5$ (and also $\partial_r\Phi_Q \neq \partial_r\Phi$). For initial $k_\delta /k_{J,{\rm MRE}} \gtrsim 2$ the average density profile of the objects that contain stars, which interpolates from the soliton to power law form, has a universal shape independent of $k_\delta$ with a power law at $r\gtrsim \lambda_J(\rho_s)$. 
Meanwhile, for initial $k_\delta/k_{J,{\rm MRE}} \lesssim 2$ the average density profile switches to a power law at smaller $r/\lambda_J(\rho_s)$ for smaller $k_\delta$. We also note that the central density of a given axion star oscillates with time by as much as an order of magnitude, indicating that the stars are produced with quasi-normal modes excited. 
Simulations of axion stars in flat space-time, which can be run for arbitrarily long times, suggest that the longest-lived quasi-normal modes persist for at least $10^2$ oscillations (detailed analysis of these modes can be found in refs.~\cite{Guzman:2018bmo,Chan:2023crj}).

We define the mass of an axion star to be the mass within the region in which the spherically averaged density profile matches the axion star prediction to within a factor of $2$ (given the criteria for identifying a star, this is inevitably not far from the mass of an isolated star with the same central density). The inferred mass of a particular axion star varies by up to a factor of $2$ throughout a single oscillation of its central density. 
In flat space-time simulations the mass of axion stars increase slowly with time due to accretion even as the quasi-normal modes decay away, 
and we anticipate the same to be the case beyond the range of cosmological simulations. 
It is less clear how the mass contained within a fuzzy-halo should be defined. To give an indication, we take the edge of the halo to be the radius at which the spherically averaged density profile drops to $20$ times the mean dark matter density, although this introduces an artificial time-dependence due to the mean density decreasing. 
\begin{figure}[t]
    \centering
    \includegraphics[width=0.95\textwidth]{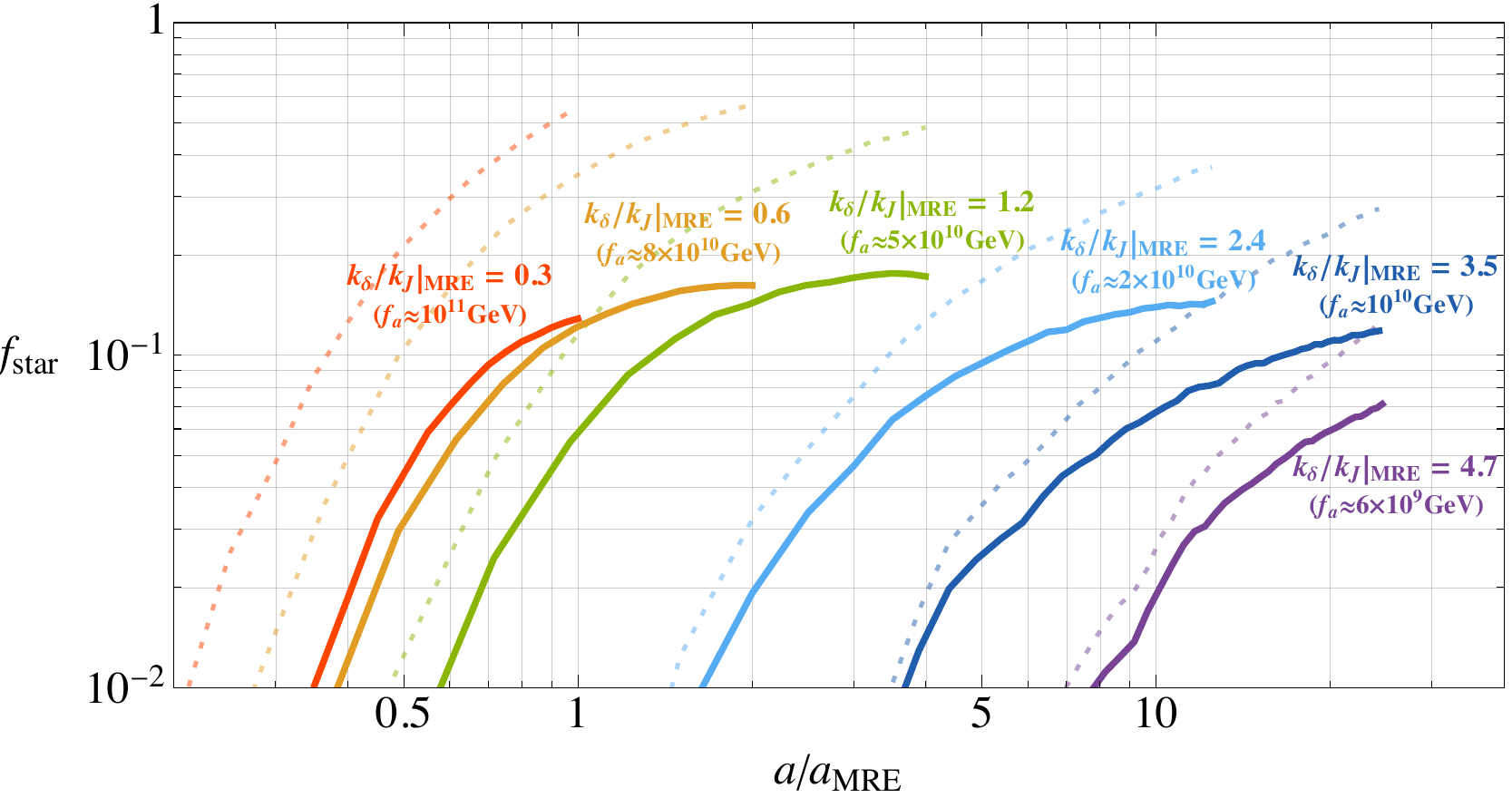}
    \caption{The fraction of dark matter axions bound in axion stars, $f_{\rm star}$, as a function of scale factor $a$ relative to that at matter-radiation equality $a_{\rm MRE}$ (solid lines). Results are shown for initial conditions with the density power spectrum peaked at different values $k_\delta$ relative to the quantum Jeans scale at matter-radiation equality $k_{J,{\rm MRE}}$. The values of the axion decay constant $f_a$ that each initial condition corresponds to are also given. Additionally, we plot the total fraction of dark matter axions in gravitationally bound objects (dashed lines), which includes those in axion stars as well as those bound in ``fuzzy halos'' around axion stars or in miniclusters that do not contain an axion star.}
    \label{fig:fstar}
\end{figure}

In Figure~\ref{fig:fstar} we plot the fraction of dark matter bound in axion stars  $f_{\rm star}$ for different initial conditions. These results are averaged over of order $10$ simulations per $k_{\delta}$, leading to statistical uncertainties of less than $10\%$. For $0.6 \leq \left. k_\delta/k_{J} \right|_{\rm MRE} \leq 2.4$ there is a burst of axion star formation  around MRE, and by the end of the simulations $f_{\rm star}$ reaches almost constant values in the range $[0.15,0.2]$, approximately independent of the initial $k_{\delta}$. This is because the majority of the order-one fluctuations have collapsed into axion stars, 
and indeed the rate at which new axion stars form decreases towards the end of these simulations, see Appendix~\ref{app:sim_mre}. There are expected to be some additional axions stars produced at later times from the collapse of fluctuations on smaller comoving spatial scales due to the increase in the comoving quantum Jeans scale, but these have progressively smaller masses $M_s \sim \bar{\rho}k_J^{-3}\propto a_{\rm coll}^{-3/4}$. 
For larger initial $ \left. k_\delta/k_{J} \right|_{\rm MRE}$ quantum pressure delays the collapse of overdensities of size $k_\delta^{-1}$ until later times. With initial $ \left. k_\delta/k_{J} \right|_{\rm MRE} \gtrsim 3$, $f_{\rm star}$ is still increasing at the end of simulations, because not all of the order-one overdensities have collapsed by this time (the rate of production of axion stars also shows no sign of decreasing).  
We expect that for $ \left. k_\delta/k_{J} \right|_{\rm MRE} \simeq 5$, $f_{\rm star}\gtrsim 0.1$ will be reached beyond the final simulation time, although we cannot determine if the asymptotic $f_{\rm star}$ is the same as for $k_\delta\simeq k_{J,{\rm MRE}}$. Finally, for initial $ \left. k_\delta/k_{J} \right|_{\rm MRE} \simeq 0.3$,  $f_{\rm star}$ reaches values larger than $0.1$ in simulations. 
In Figure~\ref{fig:fstar} we also show the total fraction of dark matter bound in axion stars, fuzzy-halos or miniclusters, which reaches values greater than $0.1$ for all initial conditions tested. 
Consistent with our analysis of the density profiles of clumps, the ratio of mass in stars to the total bound mass is similar for all initial $\kdkJ\gtrsim 2$ and is smaller for smaller $\kdkJ$. This suggests that $f_{\rm star}$ for $ \left. k_\delta/k_{J} \right|_{\rm MRE} \simeq 0.3$ might saturate at slightly smaller values than the other initial conditions. 

In combination with the results of Section~\ref{sec:new}, the initial value of $\kdkJ$ can be related to a corresponding approximate $f_a$, which we indicate on Figure~\ref{fig:fstar}.  
Remarkably, $f_{\rm star}\gtrsim 0.1$ is expected over the full range of plausible $f_a$. For $f_a \gtrsim 5\times10^{10}$ the stars form immediately at MRE whereas for smaller $f_a$  they are produced somewhat later.

\begin{figure}[ht!]
    \centering
    \includegraphics[width=0.8\textwidth]{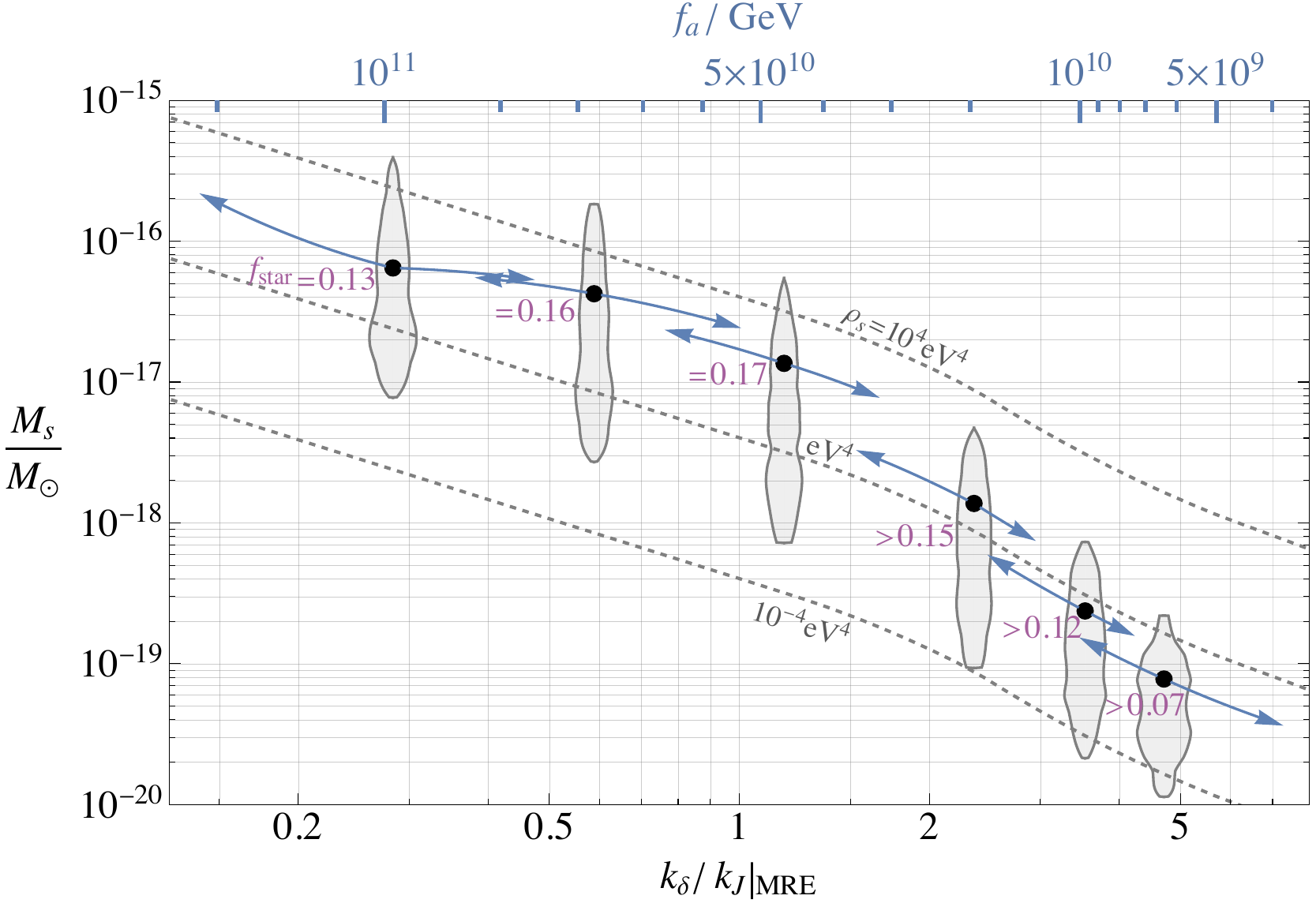}  
    \caption{The mass distribution of axion stars at the final simulation times for different initial conditions, with $k_\delta$ the peak of the initial density power spectrum. 
    The black dots show $\bar{M}_s$ defined in eq.~\eqref{eq:Mbar}; this is a weighted average that gives the typical mass of the axion stars that contain the majority of the axions that are bound in stars. The grey shapes show the distribution of axion star masses $M_s$, with the width of these regions proportional to $d\log n_s/d\log M_s$ where $n_s$ is the number density of axion stars (the area of each shape is normalised to a common, arbitrary, value). 
    On the upper axis we indicate the axion decay constants $f_a$ corresponding to the different initial conditions, with the blue error bars indicating the uncertainty in relating $f_a$ and $k_\delta$. The tilt of these curves shows the resulting uncertainty in $M_s/M_\odot$ (arising because the relation between the axion star masses and $M_\odot$ depends on $f_a$).  The dashed lines are contours of the corresponding central densities of the axion stars $\rho_s$. We also give the fraction of axions bound in axion stars at the end of the simulations, $f_{\rm star}$, which for initial $\kdkJ \gtrsim 2$ is still increasing.}
    \label{fig:vfa}
\end{figure}

The mass distribution of the axion stars is  potentially phenomenologically important. 
In Figure~\ref{fig:vfa}   
we plot the value of
\begin{equation} \label{eq:Mbar}
    \bar{M}_s= \frac{\sum_{{\rm stars}} M_s^2}{\sum_{{\rm stars}} M_s} = \frac{\sum_{ \{ {\rm axions~in~stars} \}} M_a}{N_{a,{\rm bound}}}~,
\end{equation}
where $M_a$ is the mass of the star that an axion is bound in, and $N_{a,{\rm bound}}$ is the total number of axions in stars.  In other words, most of those axions that are in axion stars are contained in stars with mass of approximately $\bar{M}_s$. 
We also show the distribution of axion star masses, $d\log n_s/d\log M$ where $n_s$ is the number density of axion stars. 
These results are plotted for each initial $k_\delta$, and related to a corresponding $f_a$  via Figure~\ref{fig:kpfakJ}. The blue curves with arrows show the uncertainty in $f_a$, which is estimated from the upper and lower edges of the red band in Figure~\ref{fig:kpfakJ}. Numerical simulations give the axion star masses in terms of $M_{J,{\rm MRE}}$, and the relation between this and $M_\odot$, given below eq.~\eqref{eq:Mexp}, depends on the value of $f_a$. Consequently the uncertainties in $f_a$ lead to overall uncertainties in $\bar{M}_s/M_\odot$ and the mass distribution, which are indicated by the vertical displacement along the blue curves. 
Contours of fixed $\rho_s$ are also plotted, obtained from eq.~\eqref{eq:starrho} (because the mass-central density relation of the stars that form is close to that of isolated stars). 
Larger initial $k_\delta$, corresponding to smaller $f_a$, leads to stars with smaller typical central densities because these form later. 

By the end of the simulations $\bar{M}_s$ has reached an approximately constant value for initial $\kdkJ \leq 2.4$, consistent with $f_{\rm star}$ saturating,  but $\bar{M}_s$ is still decreasing for larger initial $k_\delta$, see Figure~\ref{fig:Ms} in Appendix~\ref{app:sim_mre}. 
Meanwhile, the distribution $dn/d\log M_s$ is still evolving at the final simulation time for all initial conditions, with axion stars of increasingly small mass continuing to form. 
Mergers between relatively heavy axion stars are quite rare, with only $\sim 10\%$ of the stars with mass larger than $\bar{M}_s/2$ having merged with another similarly heavy star prior to the end of the simulations (meanwhile, mergers between heavy and light stars are more common). We discuss the evolution of axion stars after formation further in Section~\ref{sec:stars}, using analytic estimates, and leave a full numerical analysis to future work.

Note that for even larger values of $k_\delta/k_J|_{\rm MRE}$ (for which
we do not have simulation results) stars will be formed even later and in smaller number,
because in this case structures at $k<k_J$ (miniclusters) will form first and wash out
smaller scale fluctuations via virialization. This is the situation for 
generic post-inflationary ALP dark matter (for which $k_p(T_\star)\gtrsim 10 H_\star$) and also for QCD axions with larger values of the domain wall number 
($N_{\rm DW}$) if the emission from strings dominates, which leads to axion decay constants at least a factor 
of $N_{\rm DW}$ smaller than for the $N_{\rm DW}=1$ case that we focus on~\cite{Gorghetto:2020qws}.

Finally, we discuss the caveats and uncertainties associated to our results. We have checked that systematic uncertainties from the lattice resolution and finite time-step are negligible compared to the other uncertainties.  The finite box size introduces an uncertainty of less than 20\% on $f_{\rm star}$ and $\bar{M}_s$, dominantly due to the shape of the initial $\mathcal{P}_\delta$ being slightly deformed from the infinite volume limit.  Analysis of these systematical errors is provided in Appendix~\ref{app:sys_mre}. There is also an uncertainty associated to the definition of an axion star. Demanding that a clump's density profile agrees with the axion star prediction to within a factor of $1.5$ rather than $2$ at $r=\lambda_J(\rho_s)/4$ and $r=\lambda_J(\rho_s)/2$ changes $f_{\rm star}$ and $\bar{M}_s$ by at most $25\%$. An alternative possible condition that $\partial_r\Phi$ and $\partial_r\Phi_Q$ match to within a factor $2$ at the same values of $r$ decreases $f_{\rm star}$ by at most $35\%$.  A further uncertainty comes from the shape of the initial $\mathcal{P}_\delta$. In Appendix~\ref{app:sim_mre} we show that changing the shape the peak of $\mathcal{P}_\delta$ by  order-one amounts, keeping $k_p$ fixed, alters $f_{\rm star}$ and $\bar{M}$ by roughly $25\%$, which does not affect our qualitative conclusions. 
If $\mathcal{P}_\delta$ was not proportional to $k^3$ at $k\ll k_p$ this would alter the rate of hierarchical structure formation on larger spatial scales at times beyond the reach of our simulations (discussed in the next Section) but would not affect our present results. Meanwhile, the form of $\mathcal{P}_\delta$ at $k\gg k_p$ is irrelevant because the corresponding fluctuations are always prevented from collapsing by quantum pressure.  
As mentioned, we have neglected possible non-Gaussianities in the axion field left over from the decay of the string-wall network \cite{Vaquero:2018tib} and we have also not considered possible non-Gaussianties arising from the self-interactions at $T\sim T_c$. It would be interesting to investigate whether such features in the axion field could alter the number of axion stars that form, but we leave this for future work. 
Additionally, we reiterate that there are uncertainties in relating $f_a$ to the initial $k_{p}$ and $k_\delta$ from the earlier evolution.


\section{Axion stars} \label{sec:stars}

As seen in the previous Sections, for any value of $f_a$ between $10^{10}~{\rm GeV}$ and $10^{11}~{\rm GeV}$, soon after MRE the fraction of axions contained in axion stars satisfies $0.1\lesssim f_{\rm star}\lesssim 0.2$. The actual value of $f_a$ mostly affects the time at which the stars form and their properties. For values of $f_a$ closer to $10^{11}$~GeV 
(corresponding to a suppressed production of axions from strings) at MRE the spectrum is slightly infrared compared to the Jeans scale and solitons form readily. At such a time the mean dark matter density is relatively large, so the axion stars in this case are comparatively dense and compact. For values of $f_a$ closer to $10^{10}$~GeV (associated with an enhanced production of axions from strings and favored by recent simulations~\cite{Gorghetto:2020qws,Buschmann:2021sdq,Saikawa:2024bta,Kim:2024wku}) the spectrum is slightly more ultraviolet than the Jeans scale at MRE. Most solitons therefore form somewhat later,
when the dark matter density has further redshifted, resulting in less compact stars. Because the initial fluctuations are on smaller comoving scales, the axion stars are also lighter in this case.

The results of the numerical simulations in Figure~\ref{fig:vfa} indicate that most of the axions in stars are contained in solitons with mass given by the empirical relation
\begin{equation}\label{eq:Msbar}
\bar{M}_s\approx 2\cdot 10^{-19}\, M_{\odot} \left( \frac{f_a}{10^{10}~{\rm GeV}}\right)^{\frac52}\,,
\end{equation}
approximately valid for $f_a$ between $10^{10}~{\rm GeV}$ and $10^{11}~{\rm GeV}$, with an energy density in their center 
\begin{equation} \label{eq:rhoAS}
\bar{\rho}_s \approx 0.1~{\rm eV}^4 \left(\frac{f_a}{10^{10}~{\rm GeV}}\right)^4\,,
\end{equation}
and a radius
\begin{equation}
\bar{R}_{0.1}\approx2.1\cdot 10^{6}~{\rm km}\ \left(\frac{10^{10}~{\rm GeV}}{f_a}\right)^{\frac{1}{2}}\,,
\end{equation}
at which the density is a factor of ten smaller than in the center. 
The exact dependence on $f_a$ should be taken with caution given the uncertainties in relating this to the position of the peak of the density power spectrum (see e.g. Figure~\ref{fig:vfa}). 

The axion stars therefore have a 
mass comparable to a mountain-size asteroid, but a radius a few times larger than
the Earth-Moon distance and a density more than four orders of magnitude larger than
the local dark matter density at the Sun's location. In their gravitational ground state, the axions bound in these stars
orbit with extremely low velocities
\begin{equation}
{\bar v}_b=\frac{k_J(\bar \rho_s)}{m}
\approx 6\cdot 10^{-13}
\left( \frac{f_a}{10^{10}~{\rm GeV}}\right)^\frac32\sim {\rm mm/s}\,.
\end{equation}

Axion stars with masses smaller than $\bar{M}_s$ continue to form at later times. These lighter and less dense solitons eventually dominate the number density of stars, but they remain a sub-dominant component of the dark matter energy density in stars (see e.g. Figure~\ref{fig:dfdlogM} right in Appendix~\ref{app:sim_mre}). Being less compact they might be more prone 
to tidal disruption during the subsequent evolution.

Such a large population of axion stars could have important phenomenological consequences. 
It is therefore crucial to understand whether they survive the cosmological
evolution and what their abundance and properties would be today. Given the large hierarchies of scales 
involved, tracking the full evolution from matter-radiation equality until the present day is a challenging task that merits a dedicated study and is beyond
the scope of our present work. We limit ourselves here to some educated 
estimates based on simple arguments and existing results in the literature to demonstrate
the potentially interesting implications and to motivate a more systematic and precise analysis (e.g. we neglect wave effects in destruction processes~\cite{Du:2018qor,Dandoy:2022prp}). Several aspects of this discussion are in common with that of vector dark matter stars in ref.~\cite{Gorghetto:2022sue}, while detailed studies for miniclusters can be found in \cite{Dokuchaev:2017psd,Kavanagh:2020gcy,Shen:2022ltx,DSouza:2024flu} (see also refs.~\cite{Du:2023jxh,Maseizik:2024qly} for related analysis).

Probably the most threatening processes that could deplete our primordial axion star population
are gravitational tidal disruptions among axion stars, with larger dark matter halos 
or with compact astrophysical objects. 
To estimate the importance of such events it is useful to recall that the critical distance $d_c$ for tidal disruption of a gravitationally bound object of mass $M_1$ and size $R_1$ (with escape velocity $v_1=\sqrt{2GM_1/R_1}$) off the gravitation potential of a second object of mass $M_2$ passing with relative velocity $v_r$ is given by the relation\footnote{Obtained
by matching the escape velocity of the gravitational bound object $v_1$ with the tidal velocity $v_t\simeq a_t \Delta t$ produced by the tidal acceleration $a_t=2G M_2 R_1/d_c^3$ and accumulated during the crossing time interval 
$\Delta t\simeq d_c/v_r$.}
\begin{equation}\label{eq:disrupt}
\frac{d_c^2}{R_1^2}\simeq\frac{v_1}{v_r}\frac{M_2}{M_1}\,.
\end{equation}
From this it follows that:
\begin{enumerate}
\item Two axion stars of equal masses $M_1=M_2$ can disrupt each other (i.e. $d_c \gtrsim R_1$) only if $v_r \lesssim v_1$, i.e. they are gravitationally bound to one another and they merge. This agrees with ref.~\cite{Schwabe:2016rze}, which finds from numerical simulations that solitons colliding with relative velocity less then the escape velocity merge into a larger star, while solitons colliding with higher velocities pass through each other basically unaffected.
\item When two axion stars of different mass (say $M_2>M_1$) get close enough, the heavier one is never disrupted while the lighter one is disrupted if 
$M_2\gtrsim M_1 (v_r b^2)/(v_1 R_1^2)$,  
where $b$ is the impact parameter (obtained by requiring $b\lesssim d_c$). 
Note that, as soon as larger clusters of stars form, the typical relative velocities among stars grow rapidly, hence a large hierarchy in masses is required for the less dense stars to be disrupted (typical values for $v_r/v_1$ in our galactic halo today are ${\cal O}(10^{10})$).

\item An axion star is not disrupted by other dark matter halos (such as miniclusters) if the latter are less dense than the axion star. This is indeed the typical situation in our case given that the axion stars are the first objects to form, at the locations of highest dark matter over-density. Non-solitonic halos formed at matter-radiation equality and later, while larger and more massive, are less dense. A possible caveat however is that the profiles produced during structure formation tend to develop an NFW shape with higher densities in the core, which could affect this conclusion; we discuss this further below.

\item Miniclusters are never disrupted by axion stars, passing outside them, whose mass is less than the total minicluster mass, which is also typically the case.

\item For an axion star passing in the vicinity of another compact object, such as an astrophysical star or a black hole,  eq.~\eqref{eq:disrupt} can be rewritten more conveniently using eqs.~(\ref{eq:starR}-\ref{eq:starrho}) as
\begin{equation}\label{eq:disrupt2}
\frac{d_c^2}{R_1^2}\simeq\frac{G M_2 m}{v_r}\qquad {\rm or}\qquad d_c^2\simeq \frac{\sqrt G M_2}{v_r \sqrt{\rho_s}} \,, 
\end{equation}
where $m$ is the axion mass and $\rho_s$ is the axion star's central density.
\end{enumerate}

From these considerations we deduce the following evolution. 
After the first axion stars (which contain most of the axions that are in stars) form around matter-radiation equality, hierarchical structure formation starts. Nearby solitons begin 
falling into the gravitational potential of larger and larger local overdensities, accelerating toward each other and virializing into larger and larger structures. We expect that a sub-dominant portion of the initial axion star population might merge (in particular those stars that by chance are formed close enough that their relative velocity remains small when they approach each other) while the majority virializes. At this time, possible encounters among stars become irrelevant. 
Such a picture is compatible with what is observed in numerical simulations, where just a few stars are seen merging, although the time extent of the simulations is limited. 
Merging could still remain relevant for less dense axion stars, because for them this effect switches off later, when the virial velocities of structure grows above the axion stars' mass ratios. We might therefore expect a change in the tail of the mass distribution function during hierarchical structure formation.

Because the axion stars that contain most of the dark matter (those with mass $\bar M_s$) 
are the most compact dark matter objects, they will not be disrupted by other dark matter halos.
Indeed from the previous considerations this is probably true even for most of the rest 
of the axion star population. Moreover, from the estimates above, it is plausible that at least part of the fuzzy halos hosting the axion stars might also survive the hierarchical structure formation phase.

As bigger dark matter structures form, new axion stars could be created through gravitational relaxation~\cite{Levkov:2018kau}. These solitons, appearing later in the evolution, 
are expected to grow more massive and compact than our bulk axion star population, 
although they will necessarily be much rarer. It is plausible that the two 
different populations coexist.\footnote{Note that these more dense axion
stars are not expected to pose a threat to our dominant axion star population.
Their growth rapidly slows down after their escape velocity reaches 
the virial velocity of the host halo. From our analysis, a less dense axion star 
encountering such an object is only marginally disrupted if it passes within an axion 
star radius distance, which is quite a rare event.}

During structure formation larger and larger dark matter halos develop with possibly
denser cores. A virialized axion star at distance $R$ from the center of such a halo will feel a gravitational mass 
$M(R)=\int d^3x\rho(|\mathbf{x}|<R) 
$. Using eq.~\eqref{eq:disrupt2}, 
tidal disruption will happen only for those stars whose central density satisfies $\rho_s\lesssim M(R)/R^3$, which is equivalent to $\rho_s<\rho(R)$ if the integral defining $M(R)$ is dominated by the region at $|\mathbf{x}|\simeq R$, as in the case of NFW profile (neglecting baryons, the inner NFW profiles follow the scaling $\rho(r)r=\,$const). 
At our position, $R\simeq 8.3$ kpc, in the Milky Way 
the local dark matter energy density is more than four orders of magnitude smaller than those of eq.~\eqref{eq:rhoAS}, which means
that this effect should not be a problem for average axion stars in typical dark matter halos. On the contrary, smaller mass axion stars are much less dense ($\rho_s\propto M_s^4$) 
so the low mass population of the axion star distribution could easily be affected. Similar 
conclusions follow if instead of using eq.~(\ref{eq:disrupt2}) we compare the tidal force from the halo core with the gravitational force in the axion star or if we include  the effects of baryons and black holes in the galactic bulge, at least for our galaxy.


Provided the arguments above are correct, we need only to worry about possible encounters between axion stars and astrophysical compact objects.
Here we focus on such events within the Milky Way. Consider a typical axion star with mass $\bar M_s$ gravitationally bound in the Milky Way halo. The probability for it to be tidally disrupted
by a close encounter with an astrophysical object, in particular a star, with average mass $M_\star$ during 
a crossing of the galactic disk is $P_{\times}=\sigma_t n_\star$, where
$\sigma_t$ is the cross section for tidal disruption off an astrophysical star and $n_\star$ is 
the number density of stars per unit area on the disk. The former can be estimated simply from
eq.~\eqref{eq:disrupt2} 
\begin{equation}
\sigma_t=\pi d_c^2=\pi \frac{\sqrt G M_\star}{v_r \sqrt{\,\bar \rho_s}}\,,
\end{equation}
where $v_r$ is now the virial velocity ${\cal O}(10^{-3})$.
The latter is meanwhile given by $n_\star=\Sigma/M_\star$, where $\Sigma$ is the superficial density of stars on the galactic disk. Therefore the probability of disruption for a typical axion star crossing the galactic disk in the proximity of the Sun, where $\Sigma \simeq 50\,M_\odot/$pc$^2$ \cite{Kuijken:1989hu,Flynn:2006tm}, is
\begin{equation}
P_{\times}=\pi \frac{\sqrt G \,\Sigma}{v_r \sqrt{\,\bar \rho_s}}=
6\cdot 10^{-4} \left( \frac{{\rm eV}^4}{\bar \rho_s}\right)^{\frac12} \left(\frac{\Sigma}{50\, M_\odot/{\rm pc}^2}\right)\,.
\end{equation}

This probability should be multiplied by the number of revolutions around the galaxy between formation and today.
For low eccentricity orbits the revolution time is $T_{\rm rev.}\sim 2\pi d_\odot/v_r$ 
(where $d_\odot\simeq 8.3~$kpc is the distance of the Sun from the center of the galaxy),
but for typical dark matter eccentricities in the halo, $e\simeq 0.9$ \cite{Benson:2004fb}, $T_{\rm rev.}$ can be an order of magnitude larger. 
The average number of revolutions until today is therefore $N_{\rm rev.}\simeq {\cal O}(100)(1-e)$. The fraction of axion stars crossing the galactic disk at the Sun's 
location that are tidally disrupted is therefore expected to be small with $P_{\times}N_{\rm rev.}\sim 10^{-2} \div 10^{-1}$. 

We also note that the galactic dark matter halo extends much further from the center of the galaxy than the baryonic disk does, so 
most of the axion stars have larger orbits, with even smaller disruption probabilities, than in the estimates above. Only
the tiny fraction of axion stars with very large eccentricities, again constituting a small fraction of the population, will pass close enough to the center of 
the galaxy (where the number density of astrophysical objects is large) to be destroyed. 

We are therefore led to assume that most of the axion stars with mass of order $\bar{M}$ formed at MRE survive until today, maintaining similar properties,  
and in particular that this is the case for stars with orbits in the Milky Way halo comparable or larger than that of our solar system. 

The local number density of such axion stars  
is $n_s=f_{\rm star} \rho_{\rm DM}^{\rm loc}/\bar M_s$, where $\rho_{\rm DM}^{\rm loc}$ is the average local dark matter density, which we take to be $0.4~{\rm GeV}/{\rm cm}^3$. Consequently, the average distance between two axion stars is
\begin{equation}
n_s^{-1/3}=1.4\cdot 10^{8}\,{\rm km} \left( \frac{\bar M_s}{10^{-19}\,M_{\odot}} \right)^{\frac13}\left( \frac{0.1}{f_{\rm star}} \right)^{\frac13}\,,
\end{equation}
i.e. typically there are four axion stars within one astronomical unit of us and $10^5$ in the solar system at any given time! Consequently, the rate at which solitons pass through Earth might be non-negligible. 
Indeed we can estimate the average waiting time on Earth for an encounter with an axion star 
of mass $M_s$ to be
$\tau_\oplus=1/(\sigma_\oplus n_s v_r)$, where $\sigma_\oplus=\pi R^2$ is the geometric cross section for encountering an axion star at distance $\leq R$ from its center and $v_r$ the virial velocity. Substituting the relevant values we get
\begin{equation}\label{eq:tau_earth}
\tau_\oplus=5~{\rm yrs} \left(\frac{R_{0.1}}{R} \right)^2 \left(\frac{0.1}{f_{\rm star}} \right)\left(\frac{\bar{M}_s}{10^{-19}M_\odot} \right)^3\left(\frac{10^{10}~{\rm GeV}}{f_a} \right)^4 \,,
\end{equation}
which, substituting $\bar{M}_s$ from eq.~\eqref{eq:Msbar}, suggests that for low $f_a$ the axion stars' encounters with Earth might be interestingly frequent! 
Such encounters would last for an interval
\begin{equation}
\Delta t\simeq \frac{2R_{0.1}}{v_r}\sqrt{1-\frac{R^2}{R_{0.1}^2}}=8~{\rm hrs} \left(\frac{10^{-19}M_\odot}{\bar{M}_s} \right)\left(\frac{f_a}{10^{10}~{\rm GeV}} \right)^2 \sqrt{1-\frac{R^2}{R_{0.1}^2}}\,,
\end{equation}
which can be long compared to timescales relevant to experiments. 
One may wonder whether axion stars would be tidally disrupted by the Earth or the Sun before reaching the surface of the Earth. In fact an axion star passing through the Earth will be disrupted, however this would happen much after the star has left the Earth.\footnote{Indeed the dispersion $\Delta x$ accumulated by the axions in the axion stars over the interval
of time $\Delta t=d_\oplus/v_r$ that it takes for the star to pass by the solar system ($d_\oplus=1\,$AU being the Earth-Sun distance, i.e. the impact parameter with the Sun of the axion star passing through the Earth) is $\Delta x=\frac12 a_t \Delta t^2$, where $a_t \simeq G M_\odot R_{0.1}/d_\oplus^3$ is the tidal acceleration in the axion star. The relative dispersion of axions accumulated before reaching the Earth is therefore only $\Delta x/R_{0.1} \simeq \left(G M_\odot \right)\left(d_\oplus v_r^2 \right)\sim 10^{-2}$.}

Given that the dark matter density inside the axion stars could be more than four orders of magnitude larger than the average local one, for axion dark matter experiments looking in this mass range, such as those in refs.~\cite{MADMAX:2019pub,BREAD:2021tpx}, a broadband strategy might be more competitive than a resonant one. Note also that there could be many more 
stars with smaller density but bigger cross sections flying around. While their survival probability is less certain, their presence could have important implications for experiments: 
if at least part of this population survives until today it could completely alter expectations for
the local dark matter density observed on Earth, which would fluctuate continuously 
by orders of magnitudes over time scales of order days or more. The streams resulting from the destruction of such stars could also lead to interesting features in the dark matter density \cite{Tinyakov:2015cgg,OHare:2023rtm}.

\section{Conclusion}\label{sec:concl}

To summarize, we find that in the axion post-inflationary scenario the dark matter power spectrum resulting from the decay of topological defects is peaked near the quantum Jeans scale at matter-radiation equality, almost independently of the uncertainties in the precise value of the QCD axion mass that fits the observed dark matter abundance, at least provided $f_a\gtrsim 10^{10}~{\rm GeV}$
(see Figure~\ref{fig:kpfakJ}).  This coincidence enhances the number of axion stars that form around the time of matter-radiation equality. Numerical simulations confirm this expectation showing that 10 to 20\% of the total dark matter axions end up gravitationally bound in solitonic cores, which are surrounded by less dense halos containing a larger fraction of the remaining dark matter particles (see Figure~\ref{fig:fstar}). 
Estimates suggest that a sizable fraction of this primordial axion star population survives to the present day in our galaxy. The axion stars have a much larger density than the average local dark matter density in the neighborhood of the Sun (around four orders of magnitude for $f_a\simeq 10^{10}$ GeV, see eq.~\eqref{eq:rhoAS}) and one could pass though a detector on Earth every few years, see eq.~\eqref{eq:tau_earth}.

There are several important directions for future work. As mentioned, perhaps the most urgent of these is a dedicated study of the evolution of the axion stars after MRE.  To this end, it would be useful to carry out simulations from MRE until later times than we have been able to. This would require a greater separation between the box size and the lattice spacing, a challenge that is well suited to adaptive mesh refinement, as implemented in refs.~\cite{Schive:2014dra,Schwabe:2020eac}. Such simulations would allow the first stages of hierarchical structure formation, as the axion stars cluster into larger objects, to be studied and precise statistics about possible soliton disruption or mergers to be obtained. It would also be interesting to analyze the evolution of the axion stars within their surrounding fuzzy-halos, for example to determine whether the stars increase in mass due to accretion or evolve towards the core-halo relation proposed in ref.~\cite{Schive:2014hza}, and to study the decay of their quasinormal modes. The dynamics of the axion stars at even later times, within the much larger halos from adiabatic perturbations, and their probability of survival in the Earth's local environment are also critical.

Numerous possible signals of axion clumps in the post-inflationary scenario have been proposed and analyzed in the literature, and reanalysing these in light of our results would be worthwhile. 
Gravitational signals can arise from lensing effects~\cite{Kolb:1995bu}, although in the case of femtolensing (which is the most relevant process for the axion stars themselves given the masses that we find) the sensitivity is weak due to finite source size and wave optics effects \cite{Katz:2018zrn}. Heavier ``mini-halos'' from the IR part of the density power spectrum could lead to micro-lensing events  \cite{Fairbairn:2017dmf,Fairbairn:2017sil}. Particularly promising is caustic microlensing \cite{Dai:2019lud}, which might be sensitive to axion mini-halos with masses as small as $ 10^{-15}M_\odot$. Gravitational signals of dark matter clumps inside the solar system have also recently been considered \cite{Cuadrat-Grzybowski:2024uph,Jaeckel:2020mqa,Kim:2023pkx}, including for clumps in the mass range we expect for axion stars; these are interesting given our prediction of a large number of axion stars. 
There are also possible signals arising from the solitonic nature of the axion stars (which have previously been studied assuming stars form by condensation inside miniclusters) that typically rely on the axion-photon coupling. There has been extensive work on signals from collisions between axion stars and neutron stars \cite{Tkachev:2014dpa,Iwazaki:2014wta,Raby:2016deh,Bai:2017feq,Buckley:2020fmh,Witte:2022cjj,Kouvaris:2022guf} or main sequence stars \cite{Iwazaki:2022bur}. Other ideas include axion stars converting to photons in the Milky Way \cite{Kyriazis:2023ehv},  monochromatic photon signals from collapsing axion stars \cite{Kephart:1994uy,Braaten:2016dlp}, and radio emission from axion stars \cite{Levkov:2020txo}.  Finally, we reiterate that a detailed analysis of the implications of our results for direct detection experiments would certainly be valuable.

\section*{Acknowledgements}
We thank Asimina Arvanitaki, Dmitry Levkov, John March-Russell,  and Mehrdad Mirbabayi for discussions. 
We thank GGI for hospitality during stages of this work.
We acknowledge SISSA and ICTP for granting access at the Ulysses HPC Linux Cluster, and the HPC Collaboration Agreement between both SISSA and CINECA, and ICTP and CINECA, for granting access to the Marconi Skylake partition. We also acknowledge use of the University of Liverpool Barkla HPC cluster.  EH acknowledges the UK Science and Technology Facilities Council for support through the Quantum Sensors for the Hidden Sector collaboration under the grant ST/T006145/1 and UK Research and Innovation Future Leader Fellowship MR/V024566/1.  The work of MG is supported by the Alexander von Humboldt foundation and has been partially funded by the Deutsche Forschungsgemeinschaft under Germany’s Excellence Strategy - EXC 2121 Quantum Universe - 390833306.
\appendix

\section{Details of the self-interactions}\label{app:self}

\subsection{The non-relativistic axion field} \label{aa:nonrela}

After the string-wall network collapses at $T\simeq T_\star$, the axion field $\phi$ follows the Klein--Gordon equation of motion with potential $V=\frac12m^2\phi^2 -\frac{1}{4!}\lambda\phi^4+\dots$, with temperature-dependent (i.e. time-dependent) mass $m$ and quartic coupling $\lambda$.

In the non-relativistic limit, it is convenient to rewrite the equation of motion of $\phi$ in terms of 
$\psi$ defined by eq.~\eqref{eq:phi_t} in the main text, which leads to eq.~\eqref{eq:s-tdep}.  
The gravitational potential $\Phi$ satisfies
\begin{align}\label{sp-tdep}
	\nabla^2\Phi= \frac{4\pi G}{a^3} \frac{m}{m_0} \left(|\psi|^2-\overline{|\psi|^2}\right)\, ,
\end{align}
where, as in the main text, an over-line denotes the spatial average ($m_0$ is the zero-temperature axion mass). $\Phi$ has a negligible effect in the axion field's equation of motion deep in radiation domination, including when $T\gtrsim T_c$.  
Still considering the non-relativistic limit, the axion energy density and number density are given by, respectively,
\begin{equation}\label{eq:dens_number}
	\rho=\frac12\dot{\phi}^2+\frac12m^2\phi^2=\frac{m}{m_0}\frac{|\psi|^2}{a^3} 
 \, , \qquad \qquad n=\frac{\rho}{m}=\frac{|\psi|^2}{m_0a^3} \, .
\end{equation}
The spatially averaged comoving  number density, $\bar{n} a^3 = \overline{|\psi |^2}/m_0$, is conserved even when $m$ is temperature-dependent, while the average of $\rho a^3$ is not conserved and instead increases while $m$ is growing. 

At $T\gtrsim T_c$, the axion mass is temperature-dependent. We model this dependence as 
\begin{equation}\label{eq:maTan}
\frac{m}{m_0}= \left[1+\left(\frac{T}{T_c}\right)^{\frac{\alpha }{2 r}}\right]^{-r} ~,
\end{equation}
with $\alpha\simeq 8$ and $r\simeq 0.4$ providing reasonable fits to lattice data \cite{Borsanyi:2016ksw}, where we set $T_c=155 \, {\rm MeV}$ (note that $m^2\propto T^{-\alpha}$ for $T\gg T_c$). Although the quartic coupling at $T>T_c$ is not well determined from current lattice results, we assume that it takes the form
\begin{equation}\label{eq:lambdaTan}
\lambda=
 c_0\left[1+c_0^{\frac{1}{2r}}\left(\frac{T}{T_c}\right)^{\frac{\alpha }{2 r}}\right]^{-2r}\frac{m^2_0}{f_a^2} \, ,
\end{equation}
where $c_0= 0.35$ and, as before, $\alpha\simeq 8$ and $r\simeq 0.4$. Eq.~\eqref{eq:lambdaTan} is such that $\lambda= c(T) m^2/f_a^2$ with $c(T)$ interpolating between the zero temperature value from chiral perturbation theory, $c(T\ll T_c) \simeq c_0$
~\cite{GrillidiCortona:2015jxo}, and the dilute instanton gas prediction $c(T\gg T_c)=1$.

With the preceding axion mass temperature dependence at $T\gg T_c$ (and using that the effective number of relativistic degrees of freedom at $T=T_\star$ is $g_\star\simeq 60$), we obtain
\begin{equation} \label{eq:tstarnum}
\frac{T_\star}{T_c}\simeq 20  \left(\frac{10^{10}~{\rm GeV}}{f_a} \right)^{1/6} ~.
\end{equation}
Provided that, as is expected, the string-wall network is destroyed when $m$ is roughly of order $H$, all of the strings and domain walls are destroyed soon after $T=T_\star$ with only a weak dependence on the particular value of  $m/H$ at which the destruction happens because of the fast growth of the axion mass (simulation results find that for $\log(m_r/H_\star)\simeq 6$ no strings remain after $T  \simeq T_\star/2$ \cite{Vaquero:2018tib}).

Freely-propagating modes of momentum $k_{p}$ become non-relativistic at approximately $a/a_\star= \left(\left. k_p/H_\star\right|_{T=T_\star} \right)^{1/(\alpha/2+1)}$. Therefore, in the absence of a non-linear transient,  basically all modes of interest are non-relativistic by $a/a_\star\simeq 2$ assuming $k_p(T_\star)\sim 10H_\star$. 
The situation is more complicated in the case of a sizable non-linear transient as the axion field becomes non-relativistic, as occurs for $f_a\lesssim 5\times 10^{10}~{\rm GeV}$. In this case, the temperature of the Universe when the axion field is first non-relativistic is  in the range $T_\ell=T_\star/3$ (for $f_a\simeq 10^{10}~{\rm GeV}$) and $T_\ell=T_\star/2$ (for $f_a\simeq 5\times 10^{10}~{\rm GeV}$). The axion energy spectrum is modified by such a transient, in particular its peak is shifted close to the value of the axion mass at the time when $T=T_\ell$. Full expressions for $T_\ell$ and $k_p$ after this transient, simulations confirming these dynamics, and fits of the order-one coefficients that appear in the analytic formulae can be found in ref.~\cite{Gorghetto:2020qws}. Here we simply note that (with the values of the numerical coefficients obtained in \cite{Gorghetto:2020qws}) $k_p(T_\ell)$ corresponds to
\begin{equation} \label{eq:kpkJ1}
\left. \frac{k_{p}}{k_J}\right|_{\rm MRE}  \simeq 0.4 \left(\frac{10^{10}~{\rm GeV}}{f_a} \right)^{1/2} ~,
\end{equation}
which is the red line in Figure~\ref{fig:kpfakJ}. For sufficiently large $f_a$ such a non-linear transient is absent; the simulations results in  \cite{Gorghetto:2020qws} show that the critical value of $f_a$ above which there at most a minor effect on the axion spectrum is approximately $5\times 10^{10}~{\rm GeV}$.

We also note that, as discussed in Footnote~\ref{ft:oscillon}, some oscillons will form during the decay of the string-wall network and will persist after $T=T_\ell$, however these small objects do not affect the dynamics on the larger scales.

\subsection{Analytic analysis}\label{app:self_analytic}

As discussed in Section~\ref{sec:new}, self-interactions  drive the peak of the axion energy spectrum towards the UV. The rate at which this occurs depends on the relative size of the gradient and self-interaction energy densities in the Hamiltonian.

Neglecting gravity, in the non-relativistic regime the equation of motion of the axion field, eq.~\eqref{eq:s-tdep}, can be written in terms of its Fourier transform $\psi_{\bm k}(t)$ defined by $\psi(t,{\bm x})\!=\! \int\!\!\frac{d^3 k}{(2\pi)^3}\psi_{\bm k}(t) e^{-i {\bm k}.{\bm x}}$ as
\begin{equation} \label{eq:scatterB}
 i\partial_t \psi_{\bm k} =  \frac{k^2}{2m}\psi_{\bm k} - \frac{\lambda}{8m^3} \int \frac{d^3{p}\,d^3{q}\,d^3{l}}{(2\pi)^9} \psi_{\bm p}^* \psi_{\bm q} \psi_{\bm l} (2\pi)^3\delta^{(3)}({\bm k}+ {\bm p}-{\bm q}-{\bm l} )~,
\end{equation}
which holds regardless of the relative sizes of the gradient and self-interaction terms in the Hamiltonian.

For $(\nabla\phi)^2 \gtrsim \lambda\phi^4$ the non-relativistic axion momentum modes have dispersion relation $\omega_{\bm k} \simeq k^2/(2m)$ (up to small corrections from the self-interactions) and the system is well described by kinetic theory \cite{Semikoz:1995rd,Levkov:2018kau,Chen:2021oot,Kirkpatrick:2021wwz,Jain:2023tsr}. 
In this limit eq.~\eqref{eq:scatterB} implies (see e.g. \cite{Jain:2023ojg} for a derivation)
\begin{equation} \label{eq:dtf}
\partial_t f_{\bm k} = \frac{\lambda^2}{32 m^4} \int \frac{d^3{ p}\,d^3{q}\,d^3{ l}}{(2\pi)^9} (2\pi)^4 \delta^{(4)}(k+p-q-l) \left( (f_{\bm k}+f_{\bm p})f_{\bm q} f_{\bm l} - (f_{\bm q}+f_{\bm l})f_{\bm k} f_{\bm p} \right)  ~,
\end{equation}
where the mode occupations $f_{\bm k}\equiv  |\psi_{\bm k}|^2/(mV)$ (with $V$ the system volume) such that $\bar{\rho}=m\int d^3k/(2\pi)^3f_{\bm k}$, 
and the free dispersion relation is assumed in the energy part of the delta function. Using that the typical momentum and energy scales are $k_p$ and $k_p^2/2m$ in eq.~\eqref{eq:dtf}, and that $f_{\bm k}/(2\pi)^3\sim\bar{\rho}/(mk_p^3)$, this leads to the estimate of the thermalisation rate given in the main text
\begin{equation} \label{eq:tautherm}
\tau_{\rm therm} = \left(\frac{1}{f_{\bm k}} \frac{\partial f_{\bm k}}{\partial t} \right)^{-1} \simeq 64 \frac{m^5 k_p^2}{\lambda^2 \bar{\rho}^2}~,
\end{equation}
(note that the numerical coefficient on the right hand side of eq.~\eqref{eq:tautherm} is not sharply determined). Using the approximate relation $\lambda(T)\simeq m^2(T)/f_a^2$ and that the axion number density is $n=\rho/m$ we have 
\begin{equation} \label{eq:tHgeneral}
\tau_{\rm therm}H \simeq 64  \frac{f_a^4 k_p^2}{{n}^2 (T_\star/T)^6} \frac{m_\star}{m} \left(\frac{T}{T_\star}\right)^{-4} ~,
\end{equation}
which is valid both for the QCD axion and an ALP. 
The denominator in the first fraction is the axion number density redshifted back to $T_\star$ assuming comoving number density conservation (this will not coincide with the true axion number density at $T_\star$ if there is a first non-linear transient as the axion field becomes non-relativistic). 
For an ALP with a temperature-independent mass eq.~\eqref{eq:tHgeneral} simplifies to eq.~\eqref{eq:scatterALP} in the main text. 
As discussed in Section~\ref{sec:new}, such a particle satisfies $(\nabla\phi)^2 \gtrsim \lambda\phi^4$ for all $T<T_\star$ so $\tau_{\rm therm}$ is the relevant timescale for self-interactions, and moreover $\tau_{\rm therm} H \gtrsim 1$ for all $T<T_\star$.

Conversely, in the regime $(\nabla\phi)^2 \ll \lambda\phi^4$ the gradient term in the axion's equation of motion, eq.~\eqref{eq:s-tdep}, is irrelevant compared to the self-interactions (this is similar to the situation soon after preheating, as occurs in some theories of cosmic inflation, analyzed in ref.~\cite{Micha:2004bv}). As mentioned in the main text, in this limit the only timescale in the axion field's equation of motion is 
\begin{equation} \label{eq:tauv}
\tau_v= \frac{8m}{\lambda \phi^2} \simeq \frac{8f_a^2}{n} ~,
\end{equation}
which is therefore expected to set the time for the axion spectrum to change by an order-one amount.

For the QCD axion at $ T_c\ll T <T_\ell$ (where $T_\ell<T_\star$ is the temperature at which the axion field is non-relativistic after the first transient), we have $m\simeq m_\star (T/T_\star)^{-4}$, $\lambda\sim  m^2/f_a^2$ and the comoving axion number density is conserved. This leads to
\begin{equation} \label{eq:tauvHb}
\tau_vH\simeq \left(\frac{m_\star f_a^2}{n (T_\star/T)^3 }\right) \left(\frac{T_\star}{T}\right) ~.
\end{equation}
Using $(T_c/T_\star) \simeq (f_a/M_p)^{1/6}$, and $m_\star/m_0\simeq (f_a/M_p)^{2/3}$, $m_0\sim T_c^2/f_a$ we obtain
\begin{equation}
\tau_vH\sim \left(\frac{f_a}{M_{\rm Pl}}\right)\left(\frac{T_c}{T_{\rm MRE}}\right) \left(\frac{T_c}{T}\right) ~.
\end{equation}
Repeating this calculation including numerical factors leads to eq.~\eqref{eq:tauvH} in the main text. 
We see that the axion self-interactions are relevant when the comoving axion number density is larger than the naive contribution from domain wall decay $n(T_\star/T)^3\simeq m_\star f_a^2$, corresponding to $f_a\lesssim 5\times 10^{10}~{\rm GeV}$. This is indeed the case if the number of strings per Hubble patch $\xi\sim 10$ at $T\simeq T_\star$ 
and a scale-invariant or IR dominated emission spectrum, as is strongly suggested by simulation results. We also note that the thermalisation timescale valid instead for $(\nabla\phi)^2 \gtrsim \lambda\phi^4$ is related to $\tau_v$ by
\begin{equation}
\tau_{\rm therm} = \tau_v \left(\frac{k_p}{k_v}\right)^2~,
\end{equation}
where $k_v\equiv\sqrt{\lambda{\phi^2}}$ is the critical momentum such that $(\nabla\phi)^2 \simeq \lambda\phi^4$, as described below eq.~\eqref{eq:tauvH}. As a result, the timescale $\tau_v$ connects to $\tau_{\rm therm}$ continuously as $k_p$ increases towards $k_v$  
and the rate at which energy moves to the UV slows down for $k_p\gtrsim k_v$. Denoting comoving momenta $\tilde{k}\equiv k (a/a_\star)$,  a straightforward calculation gives 
\begin{equation}\label{eq:ktv}
\tilde{k}_v(T) \sim 700 \tilde{k}_\star \left( \frac{T_c}{T} \right)^{3/2} \left( \frac{10^{10}\,{\rm GeV}}{f_a} \right)^{5/6}~,
\end{equation}
where $k_\star\equiv H_\star$, for $T \gg T_c$ 
(this breaks down around $T\simeq T_c$ when the temperature dependence of the axion mass changes). 
Assuming that $f_a\lesssim 5\times 10^{10}~{\rm GeV}$ such that a first non-linear transient occurs, 
at $T=T_\ell \sim 5 T_c$ (when the axion field becomes non-relativistic) the peak of the axion spectrum is at $\tilde{k}_p(T_\ell)\simeq 200 \tilde{k}_\star \left(10^{10}~{\rm GeV}/f_a\right)^{5/6}$ (obtained from eq.~\eqref{eq:kpkJ1}). Therefore, at $T=T_\ell$ 
the spectrum peak is typically a factor of a few larger than $\tilde{k}_v(T_\ell)$. 
From eq.~\eqref{eq:ktv}, $\tilde{k}_v$ rapidly increases as the temperature drops below $T_\ell$ and, if the field's evolution were not affected by the self-interactions, the regime $k_p\ll k_v$ would quickly be reached (as discussed in Section~\ref{sec:new}, see also Figure~\ref{fig:energy_evo} below).

\begin{figure}[t]
    \centering
        \includegraphics[width=0.485\textwidth]{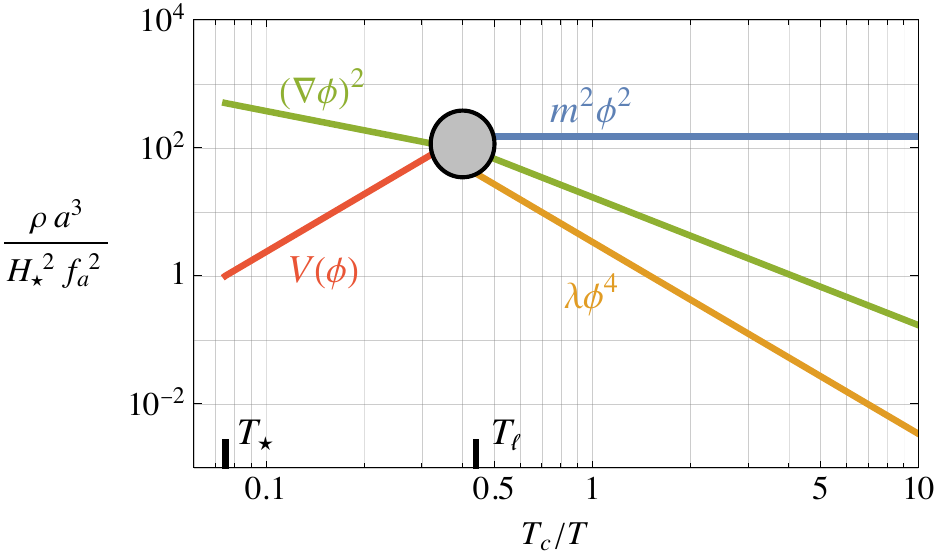}\quad
    \includegraphics[width=0.485\textwidth]{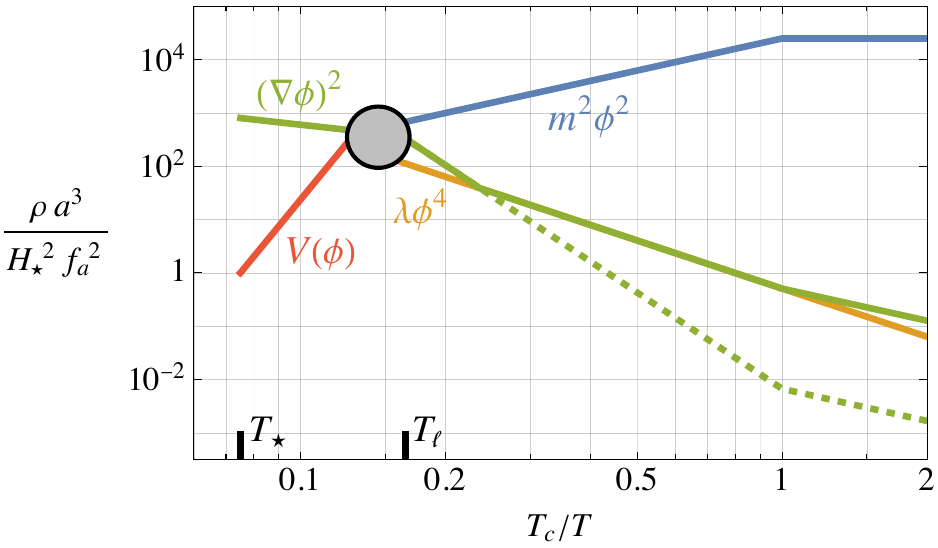}
    \caption{ {\it Left:} The schematic evolution of different contributions to the total energy density of an axion-like-particle with a temperature-independent mass as a function of the temperature of the Universe $T$ (relative to $T_c=155~{\rm MeV}$). We set the axion mass such that $H=m$ at $T=T_\star\simeq 2~{\rm GeV}$. Immediately after $T=T_\star$ the axion field is relativistic and has field amplitude larger than $f_a$. Once the potential energy $V(\phi)$ is comparable to the gradient energy, at $T=T_\ell$, there is a non-linear transient, indicated by the grey circle. Subsequently, at $T<T_\ell$, the axion field is non-relativistic with the mass contribution to the energy density dominant and the gradient energy larger than the quartic self-interaction. {\it Right:} The analogous plot for the QCD axion. There is again a non-linear transient when $(\nabla\phi)^2\simeq V(\phi)$. Subsequently, 
    the self-interaction energy $\lambda\phi^4$ decreases slower than the gradient energy $(\nabla\phi)^2$. In the absence of scattering processes mediated by the self-interactions, the quartic contribution to the energy density would exceed the gradient energy density soon after $T=T_\star$ (dashed line). However, self-interactions transfer energy to high momentum modes in such a way that $\lambda \phi^4\sim (\nabla \phi)^2$ is maintained (solid lines). This regime continues until $T\simeq T_c$ when the hierarchy $\lambda \phi^4\ll (\nabla \phi)^2$, as in the axion-like-particle case, is restored.  
    }
    \label{fig:energy_evo}
\end{figure}
The preceding expressions for $\tau_{\rm therm}$ and $\tau_v$, in combination with the scalings of the gradient and self-interaction energy densities, 
result in the effects described in Section~\ref{sec:new}. These are illustrated in Figure~\ref{fig:energy_evo}, in which we plot the evolution of the mass, gradient and self-interaction energy densities for an ALP with constant mass (left) and the QCD axion (right). Soon after $T=T_\star$, for both an ALP and a QCD axion, the field amplitude is much larger than $f_a$, the gradient energy dominates the potential energy (bounded from above as $V(\phi)\lesssim m^2 f_a^2$) and the axion evolves as a free relativistic field~\cite{Gorghetto:2020qws,Gorghetto:2021fsn}. Once the gradient energy has decreased sufficiently that $(\nabla \phi)^2\simeq V(\phi)$, the axion potential becomes relevant and there is a non-linear transient in which both $(\nabla \phi)^2$ and $V(\phi)$ change non-trivially. After this non-linear transient, the axion field is non-relativistic with amplitude less than $f_a$. 
For an ALP, the field is subsequently free and the comoving energy spectrum remains fixed until MRE. Meanwhile, 
for a QCD axion the field is free only until $(\nabla \phi)^2$ and $\lambda \phi^4$ become of the same order as a result of the rapid increase of $V(\phi)$. At this point the self-interactions become relevant and drive the peak of the axion energy spectrum close to $k_v$. 
For $f_a\lesssim {\rm few}\times 10^{10}~{\rm GeV}$ this is maintained until $T\simeq T_c$.

\begin{figure}
    \centering
    \includegraphics[width=0.48\textwidth]{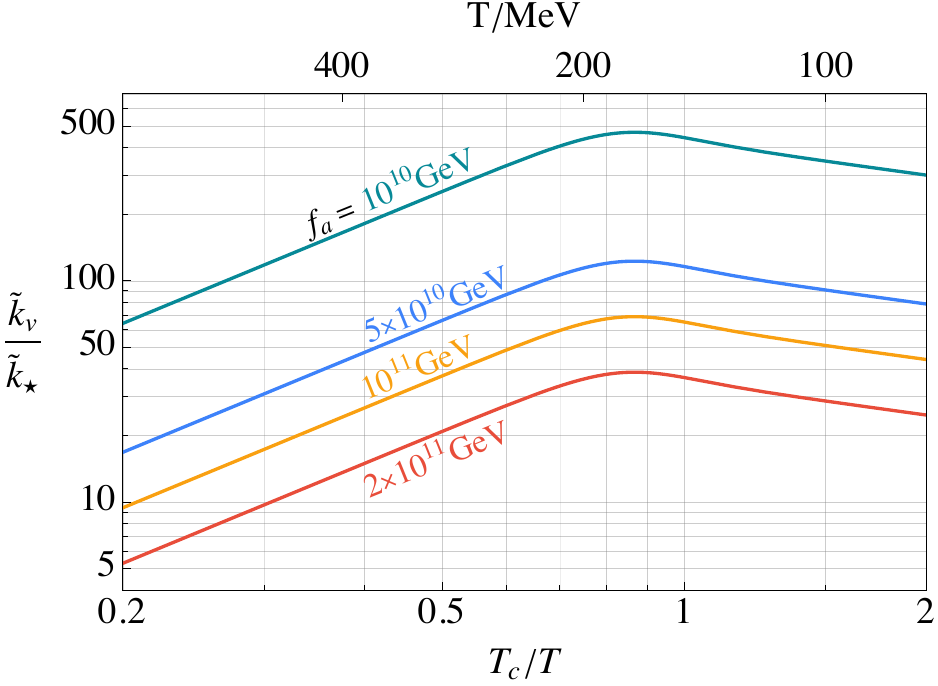}
        \includegraphics[width=0.48\textwidth]{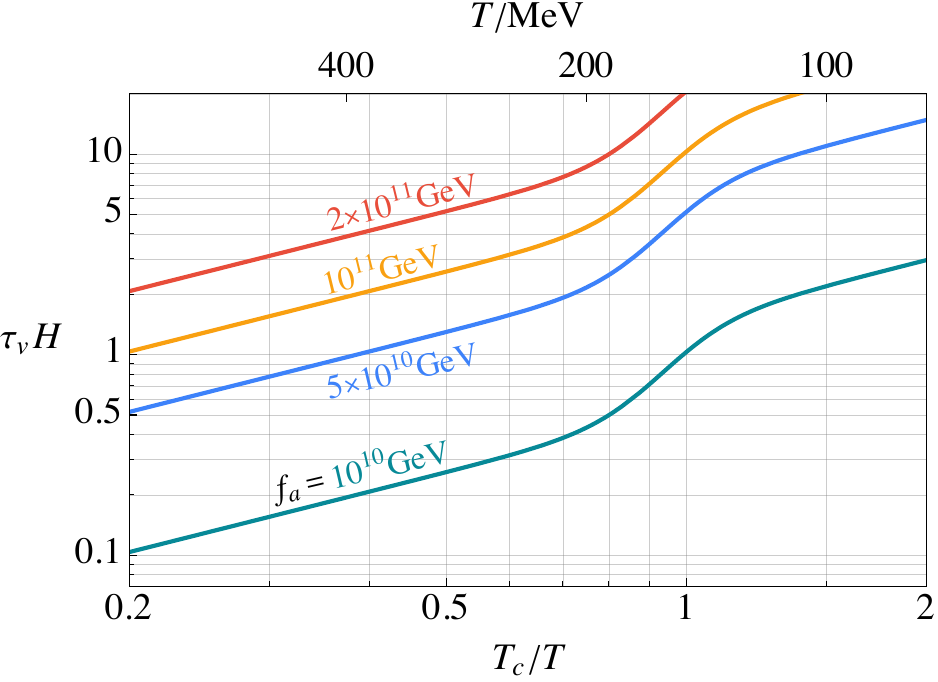}
    \caption{ \emph{Left:} The critical comoving momentum $\tilde{k}_v\equiv k_v(a/a_\star)$ such that an axion field with energy spectrum sharply peaked at 
    $k_v$ satisfies $(\nabla\phi)^2 \sim \lambda\phi^4$ (in particular such that the contributions of these resulting terms in the axion's equation of motion are equal). Results are shown as a function of the temperature of the Universe $T$ compared to $T_c=155~{\rm MeV}$, and for different values of $f_a$.  
    \emph{Right:} Comparison between the timescale $\tau_v=8m/(\lambda \phi^2)$ over which the axion energy spectrum evolves while in the regime $(\nabla\phi)^2 \ll \lambda\phi^4$  and the Hubble parameter $H(T)$. 
    }
    \label{fig:slef_pred}
\end{figure}

To see the effects of the self-interactions for a QCD axion in more detail, in Figure~\ref{fig:slef_pred} we show the temperature-evolution of $\tilde{k}_v$ and $\tau_v$ for different values of $f_a$. These are evaluated using the expressions of $m$ and $\lambda$ in eqs.~\eqref{eq:maTan} and~\eqref{eq:lambdaTan}.\footnote{Note however that we do not have control of the numerical factors in $k_v$ and $\tau_v$. For definiteness we fix the numerical factor in ${k}_v$ such that the gradient and quartic terms in the equations on motion are equal.} As the temperature of the Universe decreases from $T_\star$ to $T_c$, the critical comoving momentum $\tilde{k}_v \propto T^{-3/2}$ increases and the timescale on which self-interactions affect the spectrum increases relative to Hubble as $\tau_v H \propto T^{-1}$. We therefore expect $\tilde{k}_p$ to be roughly set by the value of $\tilde{k}_v$ when the self-interactions freeze out, i.e. when $\tau_v H\simeq 1$. As can be seen from the Figure, for $f_a\simeq 10^{10}~{\rm GeV}$, $\tau_vH\lesssim 1$ until $T\simeq T_c$ (as anticipated above) and  $\tilde{k}_v \simeq 500 \tilde{k}_\star$ at this time. Conversely, for $f_a\gtrsim {\rm few}\times 10^{10}~{\rm GeV}$ the non-linear interactions freeze out at $T\gtrsim T_c$. In the extreme case $f_a\gtrsim 10^{11}~{\rm GeV}$, $\tau_vH\lesssim 1$ only at $T\gtrsim 5 T_c$ and the corresponding $\tilde{k}_v\sim 10 \tilde{k}_\star$, so a much less dramatic change to the initial spectrum is expected.

\subsection{Setup of simulations}\label{app:simsetup}

We solve the equations of motion in eq.~\eqref{eq:s-tdep} in comoving coordinates on a discrete lattice using a second order pseudo-spectral algorithm as described in \cite{Edwards:2018ccc,Schwabe:2020eac}, see also \cite{Levkov:2018kau}. This algorithm (as with its sixth order version, which we find has similar efficiency) has the benefit of preserving the comoving axion number density $ \bar{n}a^3$ to high precision as well as being fairly fast. For simplicity, we assume that the number of degrees of freedom $g_s$ is constant for $T_c\lesssim T \lesssim T_\star$, so that $T/T_c=a(T_c)/a$. In particular, we set $g_s=25$, which is the value appropriate to temperature slightly above $T_c$ (we expect that including the full temperature dependence would have at most a minor effect on our results). 
Moreover, we assume the axion mass and quartic have the form in eqs.~\eqref{eq:maTan} and~\eqref{eq:lambdaTan}, with the parameter values quoted there (we have checked that changing $r$ by an order one factor and varying $\alpha$ in the range $[6,10]$ does not substantially affect the final position of the peak of the axion spectrum once the self-interactions freeze out at $T\lesssim T_c$).\footnote{In more detail, for $f_a\lesssim 5\times 10^{10}~{\rm GeV}$ the position of the peak is dominantly determined by the axion potential at around $T\simeq T_c$, which is not sensitive to $\alpha$ and $r$. For larger $f_a$, the values of $\alpha$ and $r$ are potentially more relevant, however $\tau_v(T)$ is approximately independent of $\alpha$ (as can be seen from the right hand side of eq.~\eqref{eq:tauv}) and $k_v(T)$ only depends on $\alpha$ as $k_v \propto T^{-\alpha/4}$ so the uncertainty is minor.}

We start cosmological simulations at the time when $T=5T_c$, which is a reasonable estimate of when the axion field is first non-relativistic for  the $f_a\in [10^{10},10^{11}]~{\rm GeV}$ that we consider. 
We have checked that our results do not change substantially (compared to the uncertainties we estimate in Figure~\ref{fig:kpfakJ}) if the initial  $T/T_c$ is varied by a factor of $2$ in either direction.\footnote{The temperature at which the axion field is first non-relativistic can be calculated in terms of $f_a$, but such precision is not needed for our purposes.} We fix the axion field in the initial conditions to have a Gaussian distribution with power spectrum 
\begin{equation} \label{eq:shape}
\mathcal{P}_{\dot{\phi}}(k)=m^2\mathcal{P}_{\phi}=\frac{\partial \rho}{\partial \log k} \propto \left(\frac{k}{k_p}\right)^4\left[1+4 \left(\frac{k}{k_p}\right)^\frac{5}{s}\right]^{-s} \ ,
\end{equation}
and for each $f_a$ we carry out simulations with different peak locations $k_p$. The parameter $s$ in the ansatz in eq.~\eqref{eq:shape} parameterizes the shape of the spectrum. For $s\simeq 4$, 
the shape is a good match to that emerging from the non-linear transient as the axion field becomes non-relativistic, as occurs for $f_a\lesssim 5\times 10^{10}~{\rm GeV}$ \cite{Gorghetto:2020qws}.\footnote{As mentioned in the main text, the form of the initial spectrum at $k< k_p$ is not reliably known, but this does not affect our results.} 
We assume that this shape is a reasonable fit to the initial spectrum also for larger $f_a$ (in which case it is directly determined by the string-wall decay). The uncertainties arising from our limited knowledge of the initial spectrum are discussed in Appendix~\ref{app:simearly}.

\begin{figure}[t]
    \centering
        \includegraphics[width=0.48\textwidth]{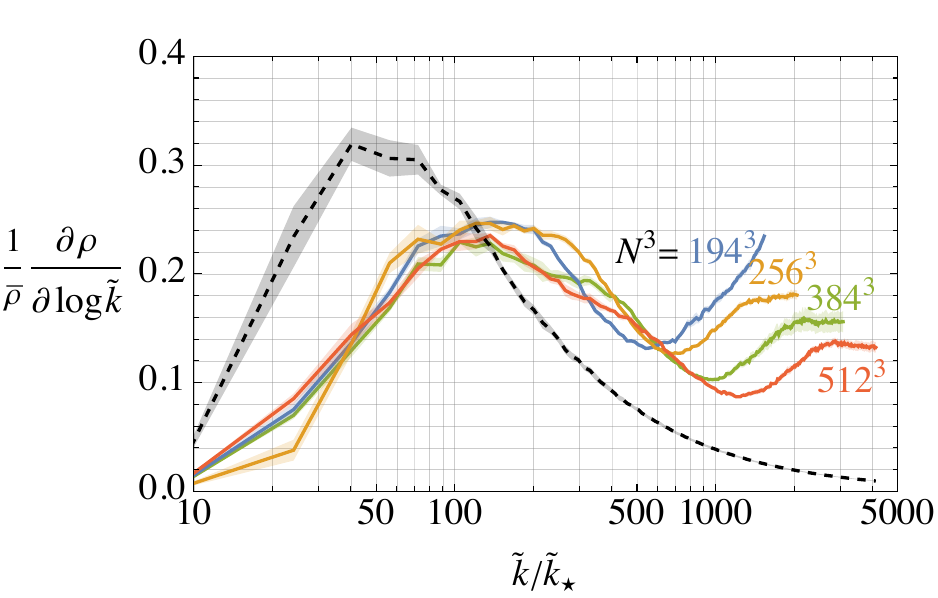}
    \includegraphics[width=0.48\textwidth]{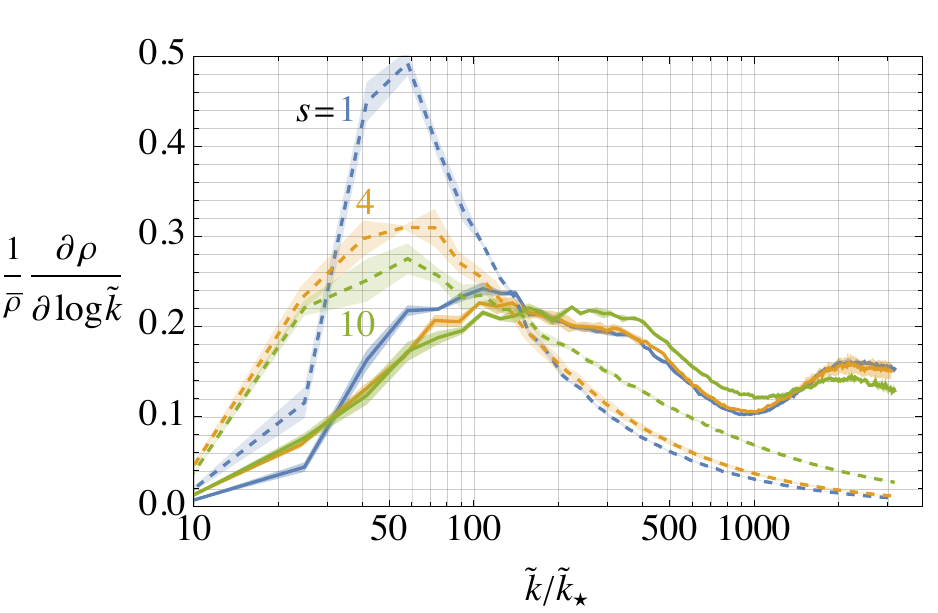}
    \caption{ \emph{Left:}      
    The comoving axion energy density spectrum at the initial (black) and final simulation times, when $T=5T_c$ and $T=T_c/2$ respectively, for $f_a=5\times10^{10}~{\rm GeV}$ and initial $\tilde{k}_{p}\simeq 50\tilde{k}_\star$. To test the impact of the finite lattice spacing systematic uncertainties, we carry out simulations with the same comoving box size $L$ and different comoving lattice spacing $L/N$ where $N^3$ is the number of lattice points. For $N^3\leq 256^3$ energy that accumulates at the lattice spacing scale affects the shape of the spectrum around its peak substantially, but for $N^3\geq 384^3$ the main part of the spectrum in approximately independent of $N$ and free of systematic errors. \emph{Right:} The axion energy spectra at the initial and final simulation times, dashed and solid lines respectively, for $f_a=5\times 10^{10}~{\rm GeV}$. Results are shown for different initial shaped spectra, parameterized by $s$ in eq.~\eqref{eq:shape}, with all spectra having initial $\tilde{k}_{p}\simeq 50\tilde{k}_\star$. Order-one changes to the initial spectrum only have a relatively minor effect on the final spectrum.}
    \label{fig:shape}
\end{figure}

Systematic uncertainties in our numerical simulations arise from the finite time-step, the finite box size, and the finite lattice spacing. We have tested the impact of these for each $f_a$ and initial $k_p$ (different simulation parameters are required to avoid systematic uncertainties in each case). The time-step is chosen sufficiently small that it introduces at most order $\%$ level uncertainty. We fix the box size $L$ large enough that the peak of the axion energy spectrum is well captured with $2\pi/L\lesssim k_p/3$ (we have tested that the results are unchanged for bigger $L$). The number of lattice points  that we have sufficient computing resources to evolve, $N^3 \lesssim 512^3$, then limits the lattice spacing. 
In Figure~\ref{fig:shape} left we plot the axion energy spectrum at the final simulation time in cosmological simulations starting from the same initial conditions for different lattice spacing. The finite lattice spacing leads to an unphysical peak in the axion energy spectrum, at momentum modes within a factor of $2$ of the lattice spacing scale, forming during the evolution. As expected, the fraction of the total energy in this unphysical peak decreases as the lattice spacing is decreased. 
If such a peak contains more than approximately $25\%$ of the total energy or is not well-separated from the main peak, the  shape and evolution of the main, physical, part of the spectrum is affected (i.e. there are significant systematic uncertainties from the finite lattice spacing). 
We therefore consider results from flat-space simulations only prior to too much energy reaching the lattice scale. Meanwhile for cosmological simulations we only consider initial values of $k_{p}$ such that  
less than approximately $25\%$ of the total energy reaches modes within a factor of 2 of the lattice spacing  
before the self-interactions freeze out and the spectrum reaches its final form at $T\lesssim T_c$.

\subsection{Results from simulations} \label{app:simearly}

\subsubsection*{Flat space-time simulations}

\begin{figure}
    \centering
    \includegraphics[width=0.495\textwidth]{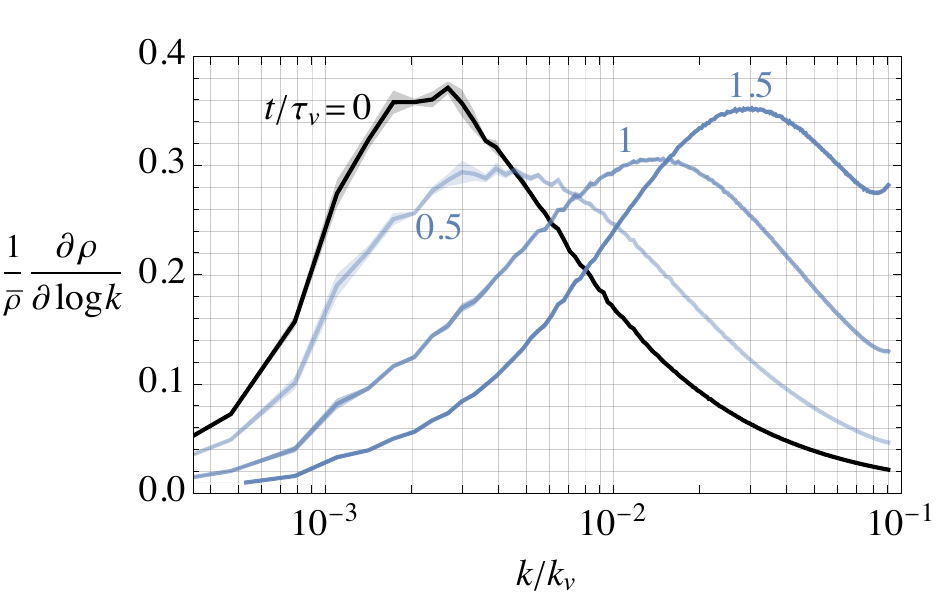}
        \includegraphics[width=0.48\textwidth]{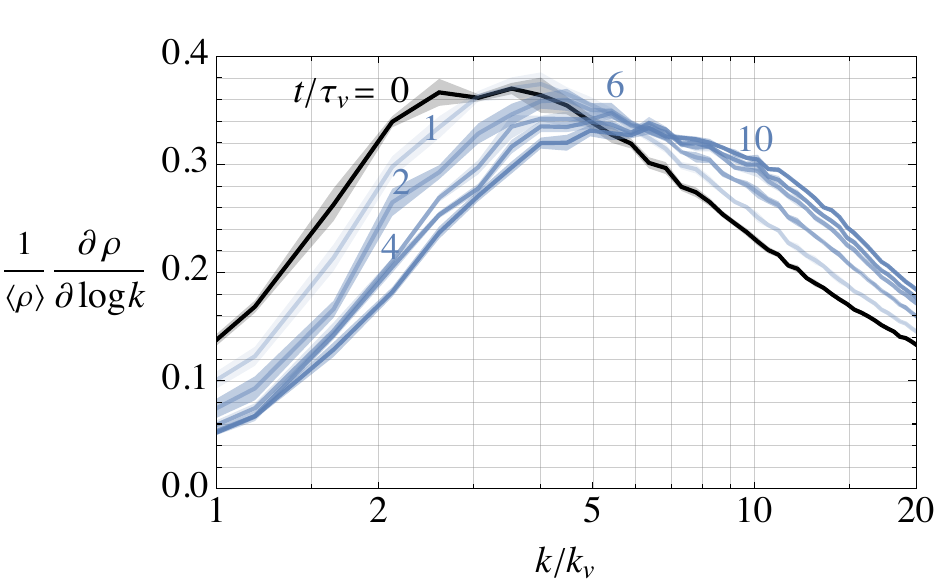}
    \caption{ \emph{Left:} The time-evolution of the axion energy density spectrum in flat space-time simulations with energy density spectrum initially peaked at  $k_p\ll k_v$. On timescales of order $\tau_v$ axion self-interactions drive the energy into the UV, to momenta much larger than $k_p(0)$.  
    \emph{Right:} Analogous to the left plot but with initial $k_p\simeq k_v$. Between $t=0$ and $t\simeq 5 \tau_v$ the spectrum moves slightly towards the UV, and subsequently its rate of change decreases.}
    \label{fig:kikf_flat}
\end{figure}

In order to confirm the validity of $\tau_v$ in eq.~\eqref{eq:tnl}, we present  
results from simulations in flat space-time, $a=1$, with $m$ and $\lambda$ time-independent so $k_v$ and $\tau_v$ are constant (normalizing length scales to $k_v$ and time to $\tau_v$, the non-relativistic equations of motion are then independent of $\bar{|\psi|^2}$).  
In Figure~\ref{fig:kikf_flat} we show the evolution of the energy density spectrum $\frac{1}{\bar{\rho}}\frac{\partial\rho}{\partial\log k}$  as a function of time for initial $k_p\ll k_v$ (left) and for $k_p\simeq k_v$ (right); as in the main text, $\rho$ is the total axion energy density, which is dominated by the mass energy such that $\frac{\partial \rho}{\partial \log k} \propto \frac{\partial n}{\partial \log k}$ with $n$ the axion number density. 
In the case $k_p\ll k_v$, energy shifts to the UV on time-scale of order $\tau_v$, with $k_p$ increasing by a factor of $10$ by $t\simeq \tau_v$ regardless of the particular initial value of $k_p$  
(we only plot results up to $1.5 \tau_v$ because after this so much energy is at the lattice spacing scale that systematic uncertainties become significant). Conversely, for $k_p\simeq k_v$ there is only an order-one change in the spectrum between $t=0$ and $5\tau_v$. Subsequently, the spectrum is approximately constant on timescales of order $\tau_v$. This is consistent with the analytic scattering calculation, expected to be valid in this regime, which gives $\tau_{\rm therm}\simeq \tau_v (k/k_v)^2$ for $k\gtrsim k_v$.

\begin{figure}
    \centering
    \includegraphics[width=0.6\textwidth]{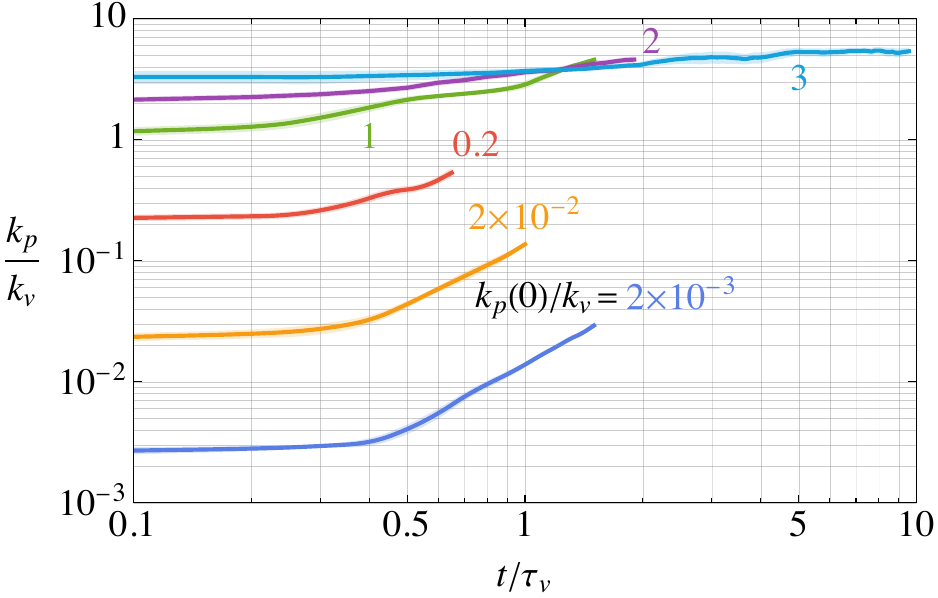}
    \caption{  The time-evolution of the peak position $k_p$ of the axion energy density spectrum $\partial\rho/\partial\log k$ in flat space-time simulations, starting from different initial $k_{p}$ (labelled $k_{p}(0)$). Momenta are normalized relative to the critical momentum $k_v$ such that $(\nabla\phi)^2=\lambda\phi^4$ and times are normalized relative to $\tau_v=8m/(\lambda \phi^2)$.}
    \label{fig:kp_flat}
\end{figure}

As further confirmation, in Figure~\ref{fig:kp_flat} we plot the time-evolution of the spectrum peak $k_p$ for different initial $k_p$, again in flat space-time simulations. As expected, while $k_p\ll k_v$ the axion spectrum moves towards the UV on timescales of order $\tau_v$. 
(For initial $k_p\ll k_v$, between $t=0$ and $t=\tau_v/2$ the shape of the spectrum changes without the location of the peak moving too much, which accounts for the plateaus in the plot at these times.)  
Meanwhile, for initial $k_p\gtrsim 3 k_v$ the spectrum peak moves at a much slower rate.

\subsubsection*{Cosmological simulations}

\begin{figure}[t]
    \centering
    \includegraphics[width=0.49\textwidth]{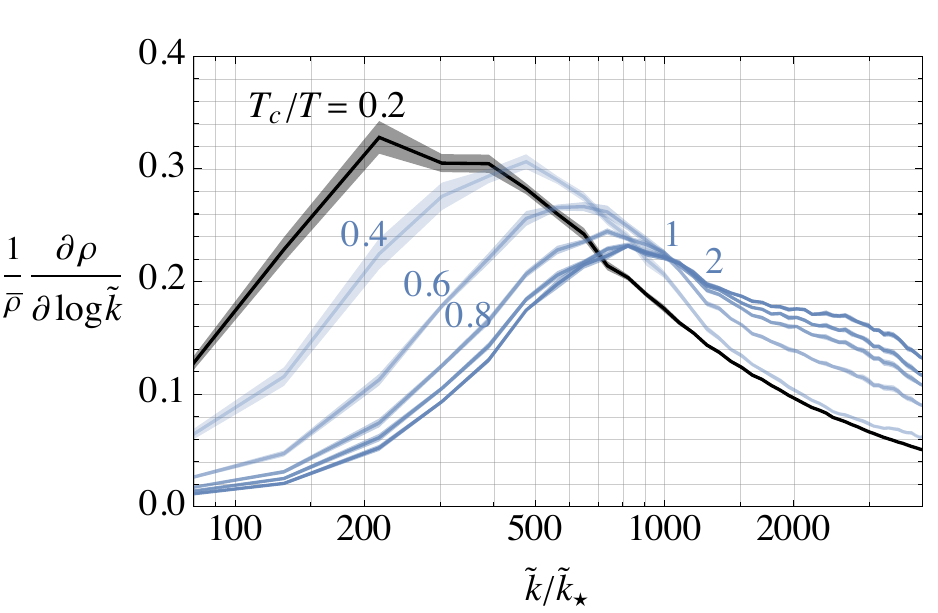}
      \includegraphics[width=0.49\textwidth]{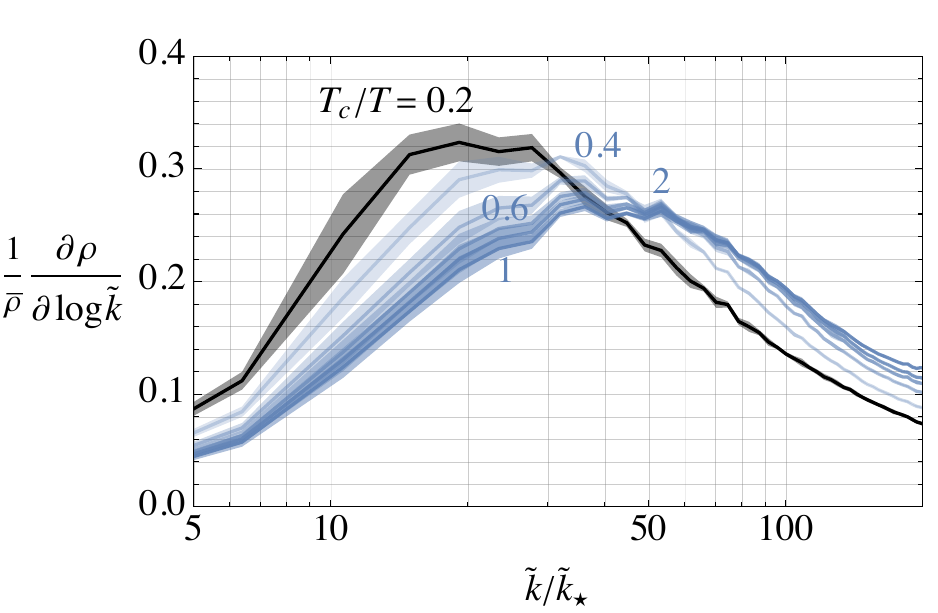}
    \caption{ \emph{Left:} The evolution of the axion energy spectrum due to self-interactions between $T=5T_c$ (roughly when the field is first non-relativistic) and $T/T_c=1/2$ (after which spectrum is close to constant), for $f_a=10^{10}\,{\rm GeV}$. The initial conditions have comoving $\tilde{k}_p\simeq 200 \tilde{k}_{\star}$, as expected after the preceding non-linear relativistic transient. The self-interactions have a substantial effect, driving $\tilde{k}_p$ towards the UV. \emph{Right:} Analogous plot for $f_a=10^{11}~{\rm GeV}$ with initial $\tilde{k}_p\simeq 20\tilde{k}_\star$. In this case the axion self-interactions have a less dramatic impact.}
    \label{fig:drhodlogk}
\end{figure}

Cosmological simulations 
of the axion field from from $T\simeq T_\ell$ to $T\lesssim T_c$ show that the energy spectrum evolves in the way expected from the analytic analysis and the flat space-time results. First, in Figure~\ref{fig:drhodlogk} we plot the energy spectrum at different $T$ for $f_a=10^{10}~{\rm GeV}$ and $f_a=10^{11}~{\rm GeV}$, with initial $k_p$ set to the physically expected values (from the preceding relativistic transient and from string decay for the two $f_a$, respectively). 
For $f_a=10^{10}~{\rm GeV}$ the axion self-interactions have a substantial effect between $T=5T_c$ and $T\simeq T_c$ increasing $\tilde{k}_p\equiv k_p (a/a_\star)$ by about a factor of $4$ and broadening the peak. The final $\tilde{k}_p$, once the self-interactions have frozen out, is a factor of a few larger than a naive estimate from the analytic results in Figure~\ref{fig:slef_pred}. This is consistent with the flat space results of Figure~\ref{fig:kp_flat} in which $k_p$ changes by order-one amounts in times $\tau_v$ for $k_p\lesssim 5k_v$. Meanwhile for $f_a=10^{11}~{\rm GeV}$ the effect of the self-interactions is somewhat smaller with only, roughly, a factor of $2$ change in $\tilde{k}_p$. In this case, the change in the spectrum mostly takes place while $T\gtrsim 2T_c$ (which is consistent with our analytic expectation in Figure~\ref{fig:slef_pred}).

\begin{figure}[t]
    \centering
    \includegraphics[width=0.7\textwidth]{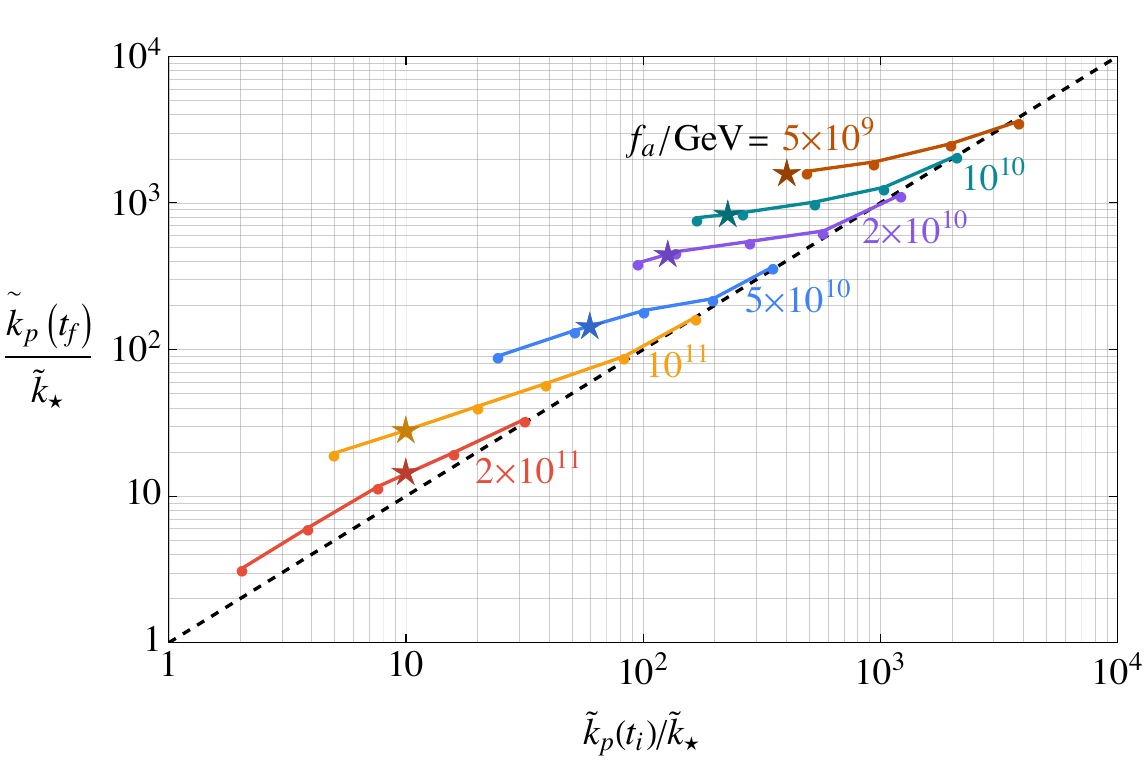}
    \caption{  
    The final position of the peak of the comoving axion energy spectrum, $\tilde{k}_{p}(t_f)$, once the axion self-interactions have frozen out (when $T<T_c$) as a function of its initial position, $\tilde{k}_{p}(t_i)$ for different $f_a$. The approximate $\tilde{k}_{p}(t_i)$ expected from the preceding dynamics for each $f_a$ is indicated with stars. The resulting values of $\tilde{k}_p(t_f)$ are used to obtain the predictions for $k_p/k_J|_{\rm MRE}$ in Figure~\ref{fig:kpfakJ} of the main text, while the uncertainties in that Figure are obtained by varying the initial $k_p$ over the range expected from the previous evolution (for $f_a\lesssim 10^{10}~{\rm GeV}$ this requires a small extrapolation in $\tilde{k}_p(t_i)$). 
}
    \label{fig:kikf}
\end{figure}

To obtain more general predictions, we carry out simulations with a range of initial $\tilde{k}_p$ for different $f_a$. 
The results are summarized in Figure~\ref{fig:kikf}, 
in which we plot the final value of $\tilde{k}_p$ as a function of the initial $\tilde{k}_{p}$ for different $f_a$. 
In that plot we also highlight the results corresponding to the physically expected initial $\tilde{k}_{p}$ (which, with our available computing resources, we can simulate for all $f_a \gtrsim 5\times 10^9~{\rm GeV}$ without significant systematic uncertainties). 
For $f_a\lesssim 5\times 10^{10}~{\rm GeV}$ the  attractor-like behaviour is evident with the final $\tilde{k}_{p}$ approximately independent of the initial conditions provided these have $\tilde{k}_{p}\lesssim \tilde{k}_v(T_c)$. For larger initial $\tilde{k}_{p}$, there is only a slight drift of the peak towards the UV. 
Meanwhile, for $f_a\gtrsim 10^{11}~{\rm GeV}$ $\tilde{k}_p$ only changes by a factor of a few regardless of the initial conditions because $\tau_v H\gtrsim 1$ at all $T<T_\star$.

Finally, to give an indication of the dependence of the final spectrum on the shape of the initial spectrum, in Figure~\ref{fig:shape} right we plot results from simulations starting from initial conditions with the same $\tilde{k}_p$ but different parameters $s$ in the ansatz in eq.~\eqref{eq:shape}. 
We fix $f_a=5\times 10^{10}~{\rm GeV}$ and initial $\tilde{k}_{p}= 50 \tilde{k}_\star$, for which the axion self-interactions have a substantial effect (the results are similar for other choices). Varying the initial shape by order-one amounts, by changing $s$ from $1$ to $10$, only changes the final $\tilde{k}_p$ by less than approximately $25\%$. Moreover, the overall shape of the final spectra are similar and almost independent of the initial $s$, being well fitted by eq.~\eqref{eq:shape} with $s$ in the range $4$ to $6$. This partially ameliorates the uncertainty from the initial conditions.

\section{Quantum pressure and axion stars}\label{app:stars}

The relevance of the quantum Jeans scale $k_J$ in the  evolution of axion overdensities is most easily seen by transforming the Schroedinger--Poisson system to the Madelung form. This is done by defining density and velocity fields $\rho$ and ${\bm v}$ with $\psi=\sqrt{a^3\rho}e^{i\theta}$ and ${\bm v}=(am)^{-1}\nabla\theta$. The Schroedinger equation then reduces to the form of the continuity and Euler equations for a perfect fluid:
\begin{align}\label{cont}
\partial_t\rho+3H\rho+ a \nabla\cdot(\rho {\bm v})=& \ 0\\
\partial_t{\bm  v}+H{\bm  v}+({\bm v}\cdot\nabla){\bm v}=&-(\nabla\Phi+\nabla\Phi_{Q}) \label{Euler}\\
\nabla^2\Phi=&\ 4\pi G(\rho-\bar{\rho}) \ ,\label{Poisson}
\end{align}
where the `quantum' pressure potential is
\begin{equation} \label{eq:phiQ}
\Phi_{Q}\equiv-\frac{1}{2m^2}\frac{\nabla^2\sqrt{\rho}}{\sqrt{\rho}} \ .
\end{equation}
Given the minus sign in eq.~\eqref{eq:phiQ}, quantum pressure opposes gravitational collapse of overdensities, and $k_J$ is the scale at which $|\nabla\Phi|=|\nabla\Phi_Q|$. 
In axion stars $\nabla{\Phi}$ in eq.~\eqref{Euler} is fully balanced by $\nabla{\Phi_Q}$ (with the velocity terms being zero), while for a conventional halo $|\nabla \Phi_Q|\ll |\nabla{\Phi}|$. 

Axion self-interactions can modify the structure and stability of axion stars. In the presence of an attractive self-interaction $V\supset -\frac{1}{4!} c_0 m^2 \phi^4/ f_a^2$ there is a maximum stable axion star mass $M_{\rm crit}\simeq 50 (f_a M_{\rm Pl})/(m \sqrt{c_0})$ (consistent with the naive expectation from comparing the third and fourth terms in eq.~\eqref{eq:s-tdep} on the axion star solution)~\cite{Chavanis_2011}, which for the QCD axion is given by
\begin{equation}
\begin{aligned}
	M_{\rm crit}\simeq 10^{-15} M_\odot  \left(\frac{10^{-3}~{\rm eV}}{m}\right)^2 ~.
\end{aligned}
\end{equation}
The self-interactions are negligible in stars with mass $M\ll M_{\rm crit}$, which is the case for all stars that form in simulations.  This is consistent with our finding, mentioned in Section~\ref{sec:simpost}, that simulations of the evolution through MRE carried out with and without self-interactions lead to practically identical results.

\section{Simulations around matter-radiation equality}\label{app:simsmre}

Our simulations of the axion field through MRE are set up analogously to those around the time $T\simeq T_c$, except with gravitational potential included (this is straightforwardly calculated from eq.~\eqref{sp-tdep} at every time-step). The initial conditions consist of a Gaussian axion field with spectrum given by eq.~\eqref{eq:shape}, and for our main data sets we fix $s=4$.

\subsection{Tests of systematic uncertainties}\label{app:sys_mre}

Systematic uncertainties in simulations through MRE arise from: i) the finite timestep, ii) the finite lattice spacing, iii) the finite box size. Our simulations are carried out in terms of the dimensionless time $d{t'} = dt/T_0$ where $t$ is the cosmic time and $T_0=4\pi^2/(16 \pi G \bar{\rho}_{\rm MRE})^{1/2}$. 
At the final simulation time, when the scale factor $a=a_{\rm max}$, the physical lattice spacing $\Delta_x=L/N $ must be less than $\lambda_J(\rho_s)$ where $\rho_s$ is the central density of the heaviest, and therefore smallest, axion star. Meanwhile, in order that the peak of $\mathcal{P}_\delta$ is captured in the initial conditions, $L k_\delta \gg 2\pi$ at the initial simulation time, and to avoid box-sized modes being non-linear $\mathcal{P}_\delta(2\pi/L)\lesssim 1$ at the final time.
\begin{figure}[t!]
    \centering
    \includegraphics[width=0.48\textwidth]{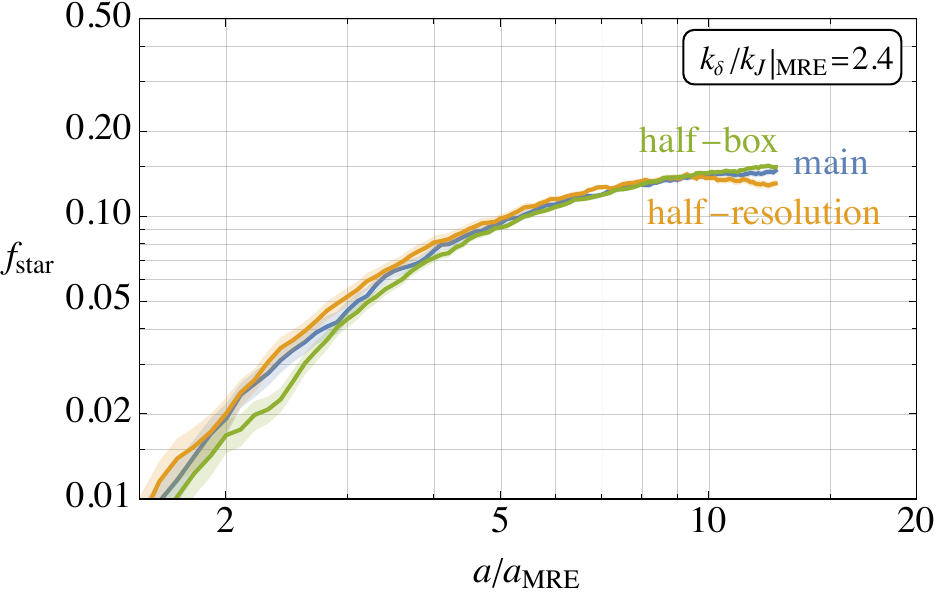}
       \includegraphics[width=0.48\textwidth]{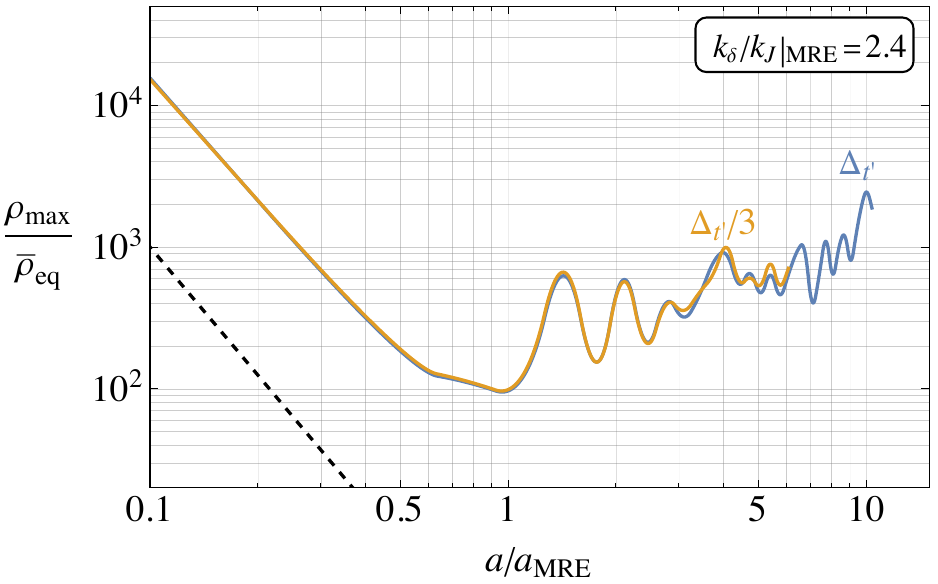}
    \caption{ \emph{Left:} The fraction of dark matter in axion stars for initial conditions with $\kdkJ =2.4$ in our main simulation runs (`main') and simulations with the comoving box size reduced by a factor of $2$ (`half-box') and with the comoving lattice spacing increased by a factor of $2$ (`half-resolution'). \emph{Right:} The maximum density obtained in simulations with initial $\kdkJ =2.4$ for simulations using the timestep of our main runs (`$\Delta_{{t'}}$') and using a timestep that is a factor of $3$ smaller (`$\Delta_{{t'}}$/3').}
    \label{fig:sys}
\end{figure}

The values of $L k_{\delta}$, $\Delta_t$, $N$ and $a_{\rm max}$ required to avoid significant systematic uncertainties  depend on the initial $k_\delta$ because this determines e.g. when overdensities collapse. We have analyzed the uncertainties for each initial $k_\delta$; here we simply report a representative example for $\kdkJ=2.4$. 
In Figure~\ref{fig:sys} left, we plot $f_{\rm star}$ for our main data set (with initial $L k_{\delta} = 24$, timestep $\Delta_{{t'}}=0.00015$ and $N^3=1152^3$) and also  with $L$ smaller by a factor of $2$ (with $L/N$ unchanged) and with comoving lattice spacing a factor of $2$ larger. 
The data obtained with smaller $L$ agrees well with the main data set.\footnote{There are slightly larger statistical uncertainties with $L/2$ because fewer axion stars form per simulation.} The results with the comoving lattice spacing twice as large agree with the main data up to $a/a_{\rm MRE}\simeq 10$ after which there is a deviation. Given that finite lattice spacing effects are expected to enter via the combination $\Delta_x/\lambda_{J}(\rho_s)$ (where 
$\lambda_{J}(\rho_s)$ is the quantum Jeans length in the core of an axion star), this suggests that our main data set is safe up to $a/a_{\rm MRE} \simeq 20$. Meanwhile in Figure~\ref{fig:sys} right we show the maximum density in simulations carried out with our main timestep $\Delta_{{t'}}$ and with the timestep a factor of $3$ smaller, with both starting from identical initial conditions. We only have sufficient computational resources to evolve the simulation with $\Delta_{{t'}}/3$ to $a/a_{\rm MRE}=6$, but up to this time the differences between the two runs are small and do not  accumulate. 

We have also checked that starting simulations at $a/a_{\rm MRE}=0.001$ or $0.1$, an order of magnitude smaller or larger than used in our main runs, does not change the results. 

\subsection{Additional uncertainties}\label{app:shape_mre}

\begin{figure}
    \centering
    \includegraphics[width=0.485\textwidth]{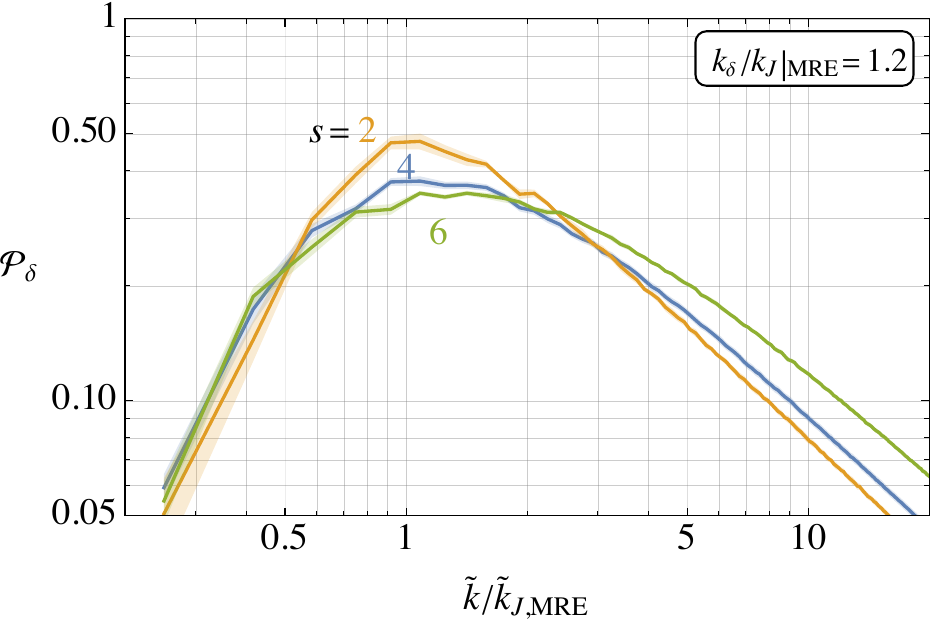}
        \includegraphics[width=0.5\textwidth]{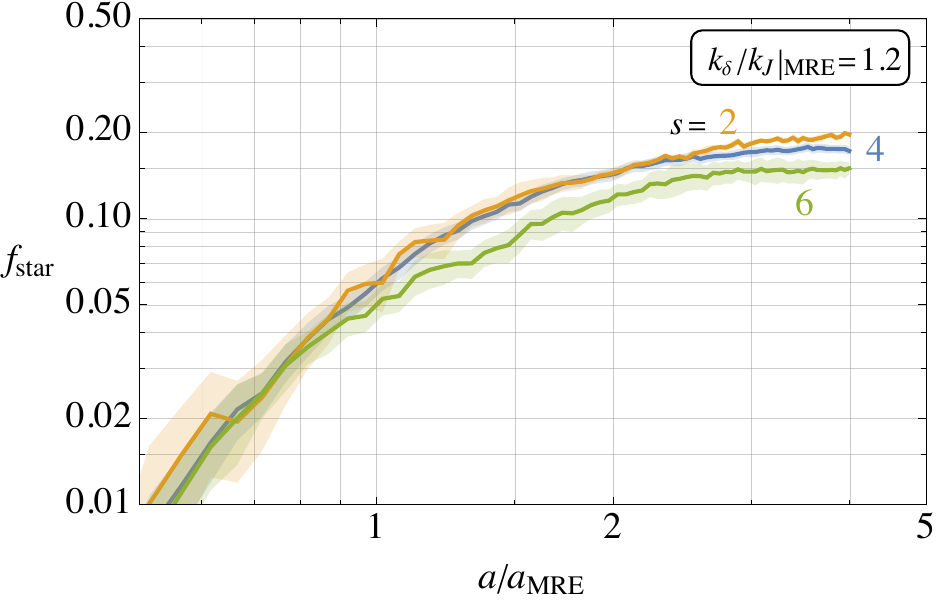}
    \caption{ \emph{Left:} The initial axion density power spectrum $\mathcal{P}_\delta$ for initial conditions with different shape parameters $s=2,4,6$ in eq.~\eqref{eq:shape}. In each case $\left. k_p/k_J\right|_{\rm MRE}=0.55$.  \emph{Right:} The fraction of axions in stars $f_{\rm star}$ in simulations starting from the initial conditions in the left panel, with statistical uncertainties.}
    \label{fig:shapeb}
\end{figure}

In our simulations around MRE we take the initial axion energy spectrum to be given by eq.~\eqref{eq:shape} with shape parameter $s=4$. To estimate the uncertainty from this assumption, we carry out simulations with initial $\left. k_p/k_J\right|_{\rm MRE} = 0.55$ and with parameters $s =2$ and $6$ (this $k_p$ corresponds to initial $\kdkJ = 1.2$ for $s=4$ and slightly different $k_\delta$ for the other $s$). In Figure~\ref{fig:shapeb} left we plot the corresponding initial $\mathcal{P}_\delta$ while in the right panel we show the resulting $f_{\rm star}$. The order-one differences in the initial $\mathcal{P}_\delta$ change the asymptotic $f_{\rm star}$ by approximately $25\%$. Meanwhile, the mass of the axion stars that the majority of the bound axions are in, $\bar{M}$, differ by approximately $10\%$ between the different $s$. 
Thus our main qualitative conclusions are robust against the details of the shape of the initial spectrum for a particular $k_p$.

\begin{figure}
    \centering
    \includegraphics[width=0.65\textwidth]{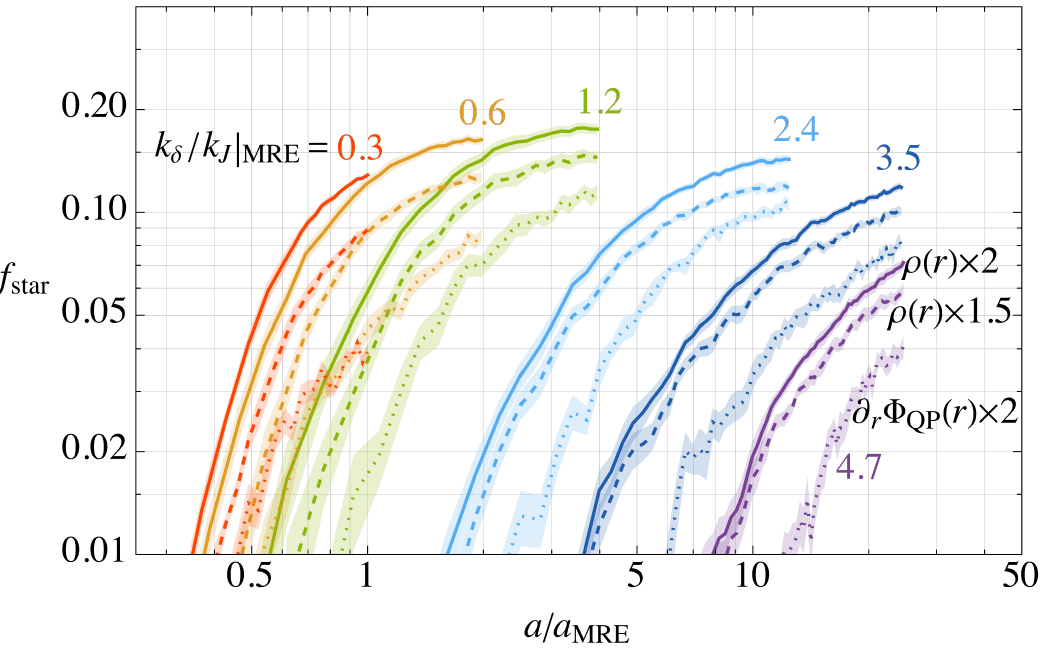}
    \caption{$f_{\rm star}$ obtained with different criteria for identifying stars: ``$\rho(r)\times 2$" by demanding that the spherically averaged density profile $\rho(r)$ agrees with the soliton prediction within a factor of $2$ at radius $\lambda_J(\rho_c)/4$ and $\lambda_J(\rho_c)/2$ (as used for our results in the main text); ``$\rho(r)\times 1.5$" similar but demanding agreement within a factor of $1.5$ at the same radius; ``$\partial_r \Phi_{\rm QP}\times 2$" requiring that the radial derivative of the spherically averaged quantum pressure $|\partial_r \Phi_{\rm QP}|$ matches the radial derivative of the gravitational potential within a factor of $2$ at the same points.}
    \label{fig:star_def}
\end{figure}

Another uncertainty comes from our criteria for classifying a collapsed object as an axion star. As discussed in Section~\ref{sec:simpost}, for our main plots we demand that an object's spherically averaged density profile matches the soliton form within a factor of $2$ at a distance $r$ of $\lambda_J(\rho_c)/4$ and $\lambda_J(\rho_c)/2$ from the star center. 
In Figure~\ref{fig:star_def}, we compare the inferred $f_{\rm star}$ with that obtained using different criteria (in particular, requiring that at the same $r$ the density profile matches the star form within a factor $1.5$ or that the radial derivative of the quantum pressure matches that of the gravitational potential to within a factor of $2$). The resulting change in $f_{\rm star}$ is typically $25\%$, and possibly as large as $50\%$ for initial $\kdkJ=0.3 $. Although not negligible, this does not significantly affect our main conclusions. We also expect that objects in which quantum pressure is relevant will evolve towards ground state axion stars at times beyond the range of simulations, e.g. as quasi-normal modes decay.

\subsection{Further results from simulations}\label{app:sim_mre}

Here we present additional results from numerical simulations of the axion field around MRE, which support our discussion in Section~\ref{sec:simpost}.

\subsubsection{Collapse of overdensities}\label{aa:collapse}

\begin{figure}
    \centering
    \includegraphics[width=0.51\textwidth]{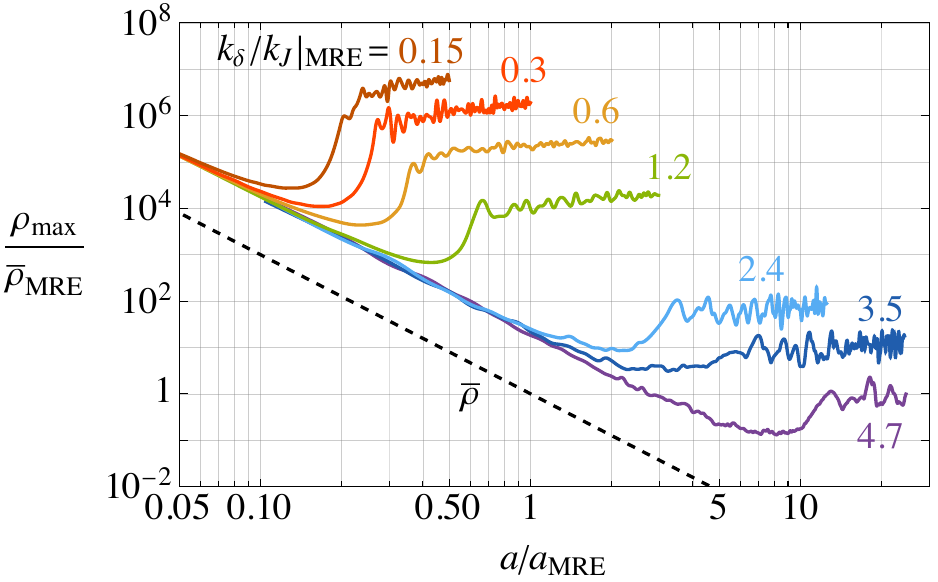}
        \includegraphics[width=0.48\textwidth]{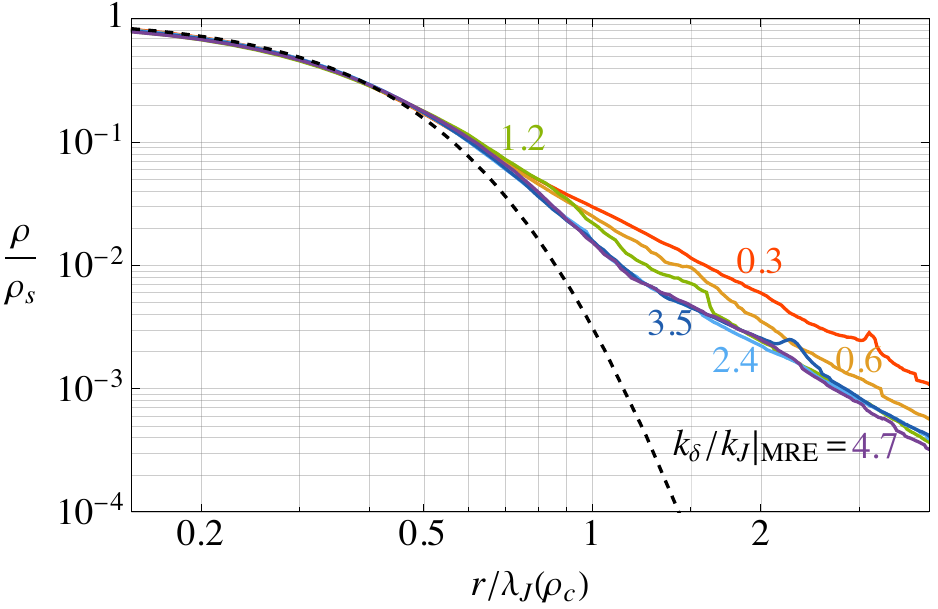}
    \caption{ \emph{Left:} The maximum density in a particular simulation $\rho_{\rm max}$ compared to the mean dark matter density at MRE $\bar{\rho}_{\rm MRE}$. The mean dark matter density $\bar{\rho}(a)$ is also plotted. The collapse of overdensities into gravitationally bound objects decoupled from the Hubble expansion is evident.   \emph{Right:} The average density profile of objects identified as axion stars for different initial conditions, normalized to the objects' central densities $\rho_s$ and the associated quantum Jeans length $\lambda_J(\rho_s)$. Results are shown at the final simulation time for each set of initial conditions. In this way, the axion star profile takes a universal form, plotted in black.}
    \label{fig:prof}
\end{figure}

In Figure~\ref{fig:prof} left we plot the maximum energy density in a single simulation run as a function of $a/a_{\rm MRE}$ for each initial $k_\delta$ (and also for initial $\kdkJ =0.15$ smaller than in our other plots, for which resolution of the collapsed objects is lost soon after formation). This is not a well-defined physical observable because the results depend on the simulation volume, but it gives useful intuition. 
As expected, quantum pressure delays the collapse of overdensities until progressively later times for larger initial $k_\delta$, playing a role even for $\kdkJ=0.3$. At collapse there is a factor of $10$ to $100$ increase in the maximum density, which is subsequently approximately constant and decoupled from the Hubble expansion. This corresponds to the center of the most dense object, with oscillations  due to the excited quasi-normal modes of the axion stars. 

In Figure~\ref{fig:prof} right we plot the mean spherically averaged density profile of all objects identified as axion stars for different initial conditions. Prior to averaging, the profile of each object is normalized to its central density $\rho_s$ and the associated quantum Jeans length $\lambda_J(\rho_s)$, such that an axion star's profile takes a universal form independent of its mass. All the averaged profiles match the axion star prediction closely for $r\lesssim 0.75 \lambda_J(\rho_s)$, but deviate at larger $r$. For initial $\kdkJ\gtrsim 2$ the averaged density profiles take a remarkably consistent form with the results for $\kdkJ =2.4,~3.5,~4.7$ basically coinciding and having a power law dependence at $r\gtrsim \lambda_J(\rho_s)$. For smaller initial $k_\delta$ the averaged density profile switches to a power law form at increasingly small $r/\lambda_J(\rho_s)$. For all initial conditions, the power law is well fit by $\rho(r)\propto r^{-2.5}$, however our results are not sufficient to distinguish between this and e.g. an NFW profile.

\begin{figure}[t]
    \centering
    \includegraphics[width=0.48\textwidth]
    {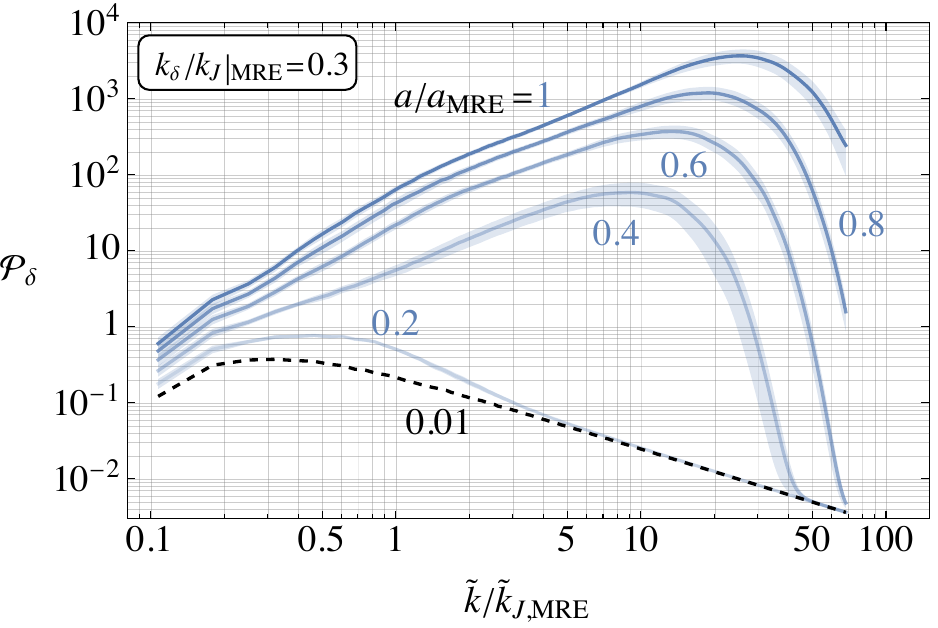}\quad
    \includegraphics[width=0.48\textwidth]{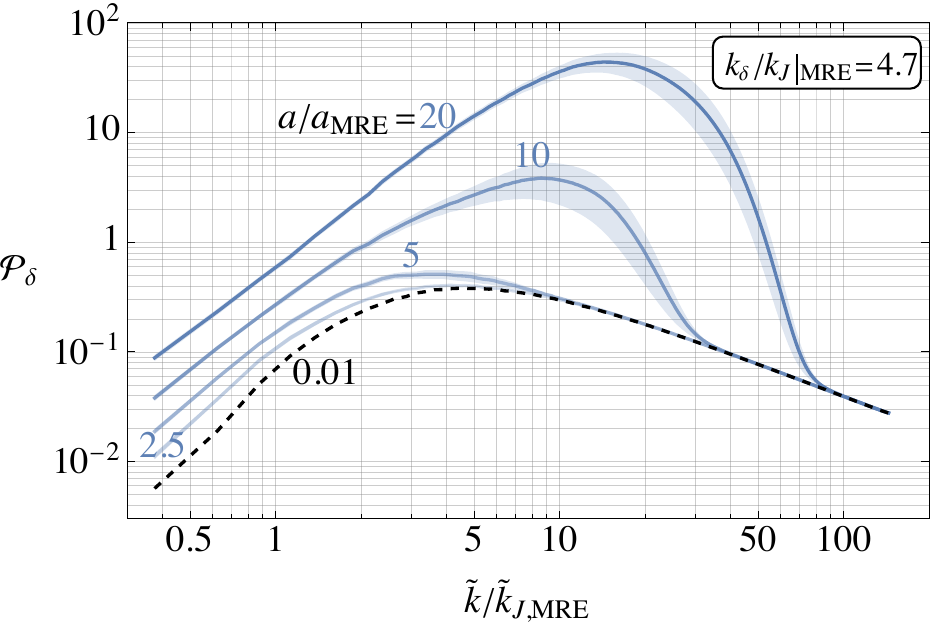}
    \caption{ \emph{Left:} The comoving axion density power spectrum $\mathcal{P}_\delta(\tilde{k})$ at different times for initial conditions with $\kdkJ=0.3$, where $\tilde{k}_{J,{\rm MRE}}$ is the comoving momentum corresponding to the quantum Jeans momentum at MRE. \emph{Right:} As in the left panel but with initial $\kdkJ=4.7$, in which case quantum pressure delays gravitational collapse of overdensities until $a/a_{\rm MRE}\simeq 5$.}
    \label{fig:Pd}
\end{figure}

In Figure~\ref{fig:Pd} we plot the comoving axion density power spectrum $\mathcal{P}_\delta(\tilde{k})$ at different times for initial conditions with $\kdkJ=0.3$ and $4.7$.  
For initial $\kdkJ =0.3$, $\mathcal{P}_\delta$ starts to increase already at $a/a_{\rm MRE}=0.2$  corresponding to collapse of overdensities evident by $a/a_{\rm MRE}=0.4$ (consistent with the results in Figure~\ref{fig:prof} left). Meanwhile, for initial $\kdkJ =4.7$ quantum pressure results in $\mathcal{P}_\delta$ remaining frozen until $a/a_{\rm MRE}\simeq 5$ at which time the comoving quantum Jeans length locally to the largest overdensities has decreased enough for them to collapse.

\subsubsection{Properties of the axion stars}\label{aa:starproperties}

\begin{figure}[t]
    \centering
    \includegraphics[width=0.48\textwidth]{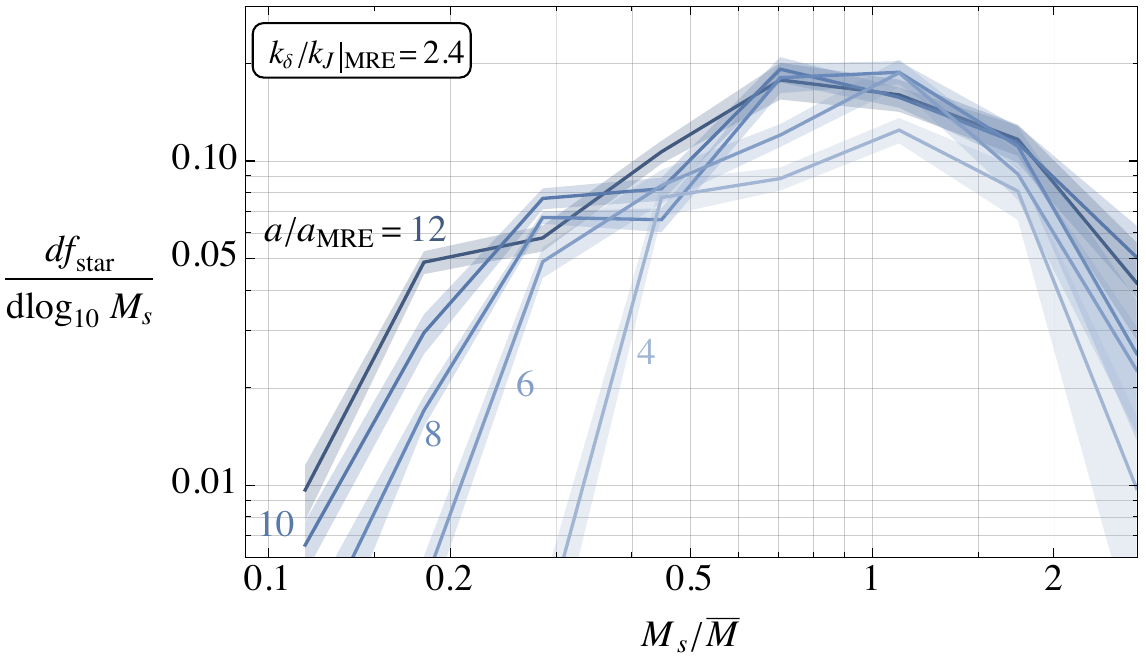}
        \includegraphics[width=0.48\textwidth]{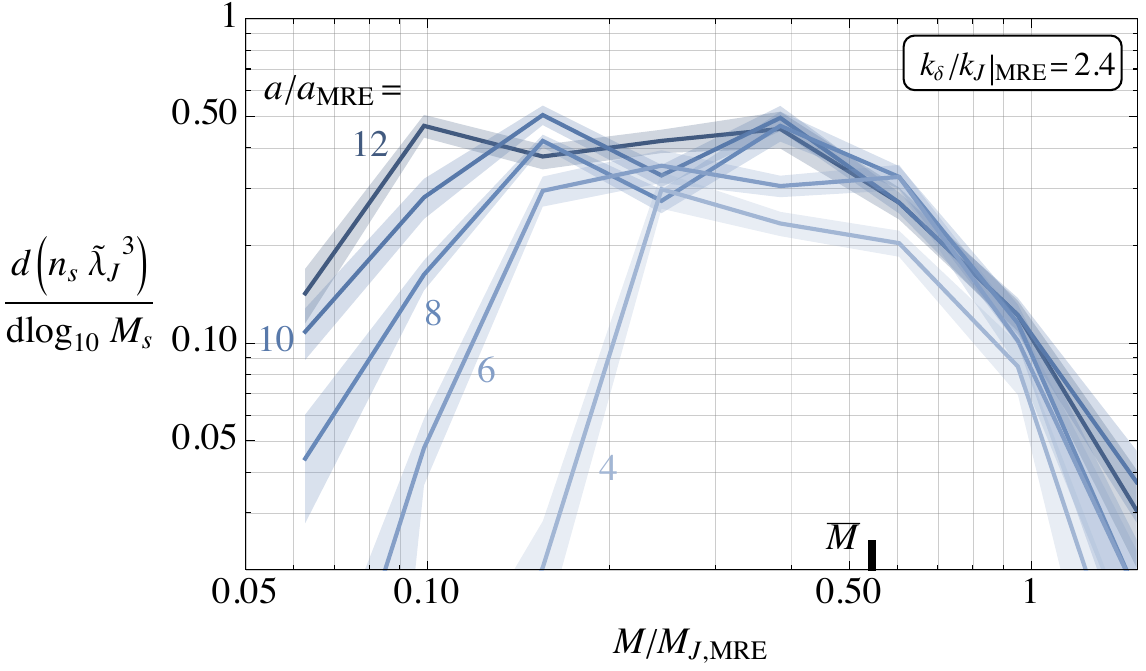}
    \caption{{\it Left:} The distribution of the fraction of dark matter bound in axion stars per logarithmic interval of axion star mass, $df_{\rm star}/d\log_{10}M_s$, at different times for initial conditions with $\kdkJ=2.4$. The bands are the statistical uncertainties on the results. {\it Right:} The distribution of the axion star number density
  $n_s$ (normalised to $\tilde{\lambda}_{J}^3 \equiv \lambda_{J,{\rm MRE}}^3 (a/a_{\rm MRE})^{-3}$, i.e. the comoving volume equal to a quantum Jeans volume at MRE) with axion star mass $M_s$. As in the left panel, results are shown at different times with initial conditions with $\kdkJ=2.4$, with statistical uncertainties. }
    \label{fig:dfdlogM}
\end{figure}

To further analyze the formation of stars,  
in Figure~\ref{fig:dfdlogM} left we plot the distribution of $f_{\rm star}$ with axion star mass: $df_{\rm star}/d\log_{10}M_s$ at different times for initial conditions with $\kdkJ  =2.4$. 
A burst of relatively heavy stars with masses $\simeq M_{J,{\rm MRE}}/2$ form at $2\lesssim a/a_{\rm MRE}\lesssim 4$ corresponding to the collapse of order-one overdensities (these make up a substantial fraction of the asymptotic $f_{\rm star}$, c.f. Figure~\ref{fig:fstar}).  Subsequently, $df_{\rm star}/d\log_{10}M_s$ is almost  time-independent at such masses. However, progressively smaller mass axion stars continue to form due to the collapse of overdensities on smaller scales as $\tilde{k}_J(\bar{\rho})$ slowly increases (these do not contribute much to the total $f_{\rm star}$). This can be seen clearly in  Figure~\ref{fig:dfdlogM} right, in which we plot the distribution of axion star number density with axion star mass. Although $f_{\rm star}$ is dominated by high masses, smaller and smaller mass stars continue to form and reach number densities comparable to those of the heavy stars.

\begin{figure}[t!]
    \centering
    \includegraphics[width=0.50\textwidth]{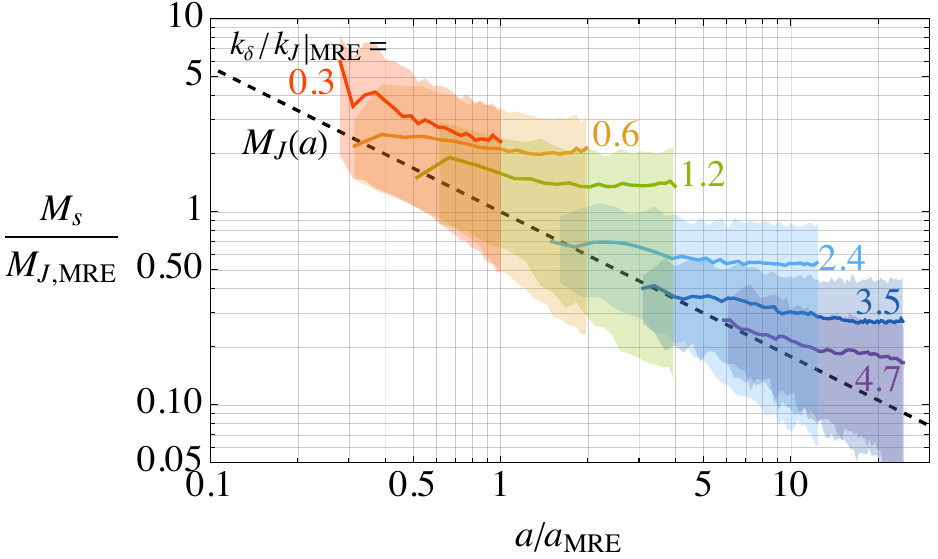}
        \includegraphics[width=0.48\textwidth]{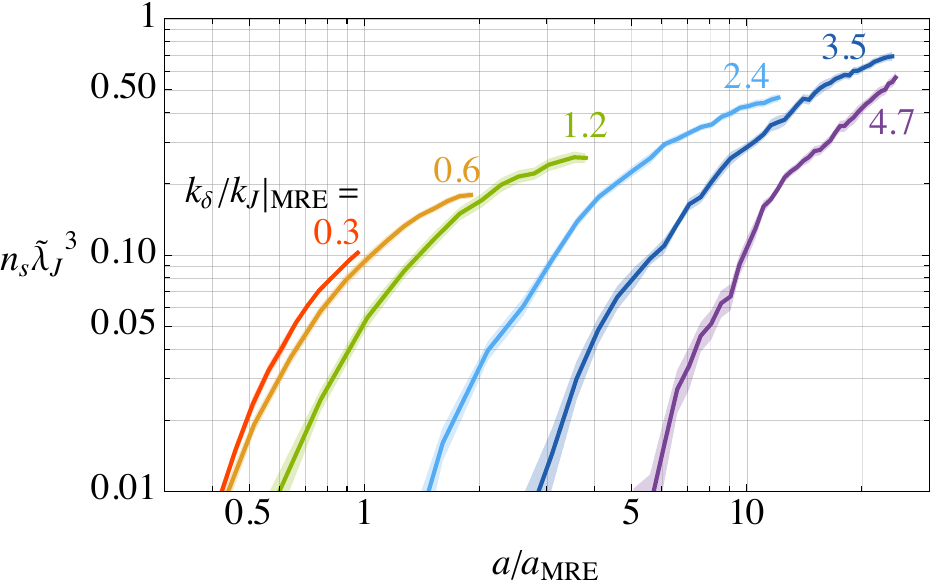}
    \caption{ \emph{Left:} The evolution of $\bar{M}_s$, defined by  eq.~\eqref{eq:Mbar}, which measures the typical mass of the axion stars  that contain the majority of bound axions. The colored bands show the interval containing $90\%$ of axion stars by number (with $5\%$ of stars are more massive and $5\%$ less massive. \emph{Right:} The number density of identified axion stars per comoving volume $n_s$ normalised to $\tilde{\lambda}_{J}^3 \equiv \lambda_{J,{\rm MRE}}^3 (a/a_{\rm MRE})^{-3}$, with statistical uncertainties.}
    \label{fig:Ms}
\end{figure}

Meanwhile, in Figure~\ref{fig:Ms} left, 
we plot $\bar{M}$, defined in eq.~\eqref{eq:Mbar}, as a function of time for different initial conditions. The colored bands in this plot indicate the distribution of axion star masses with $5\%$ of stars having mass greater than the upper edge of the bands and $5\%$ having mass smaller than the lower edge. For initial $\kdkJ \in [0.6,3.5]$, $\bar{M}$ reaches an approximately constant value by the final simulation time, which is consistent with $f_{\rm star}$ saturating. Nevertheless, the lower edges of the bands still decrease as successively lighter axion stars continue to form. For initial $\kdkJ =0.3$ resolution of the axion star cores is lost before $\bar{M}$ and $f_{\rm star}$ presumably reach constant values, while for $\kdkJ=4.7$ we do not have sufficient computational resources to evolve until $\bar{M}$ reaches a constant value at $a/a_{\rm MRE}\gtrsim 20$. 
For initial $\kdkJ \gtrsim 1$, $\bar{M}_s$ roughly matches the prediction in eq.~\eqref{eq:Mexp} with $\delta\simeq 10$. Meanwhile, for initial $\kdkJ \lesssim 1$, $\bar{M}_s$ is smaller than this estimate, which is consistent with a smaller fraction of the axions in the original overdensity being in the central axion star in this case.  

Finally, in  Figure~\ref{fig:Ms} right we plot the number of axion stars per comoving  $\lambda_{J,{\rm MRE}}^3 (a/a_{\rm MRE})^{-3}$ volume as a function of time. This continues to grow throughout the simulations for all initial conditions. The rate of increase of the number of axion stars slows towards the final simulation times for initial $0.6 \lesssim \kdkJ \lesssim 3.5$. Consistent with our analysis in Section~\ref{sec:new}, this suggests that the majority of the initially order-one overdensities on length scales of order $k_p^{-1}$ have collapsed by the end of the simulations for such initial conditions (with only a few additional, lower mass, stars continuing to form). For initial $\kdkJ = 4.7$ the number of axion stars is still increasing fast at the final simulation time, which is consistent with our expectation that the fraction of axions bound in stars $f_{\rm star}$ will continue to increase beyond the range of simulations in this case. Additionally, for $\kdkJ = 0.3$ the increase in the number of stars shows no sign of slowing before the end of the simulations. 

\clearpage

\printbibliography

\end{document}